\documentclass[aps,prl,reprint,superscriptaddress,longbibliography,floatfix]{revtex4-2}

\usepackage{amsmath,amssymb,bm}
\usepackage{graphicx}
\usepackage[dvipsnames]{xcolor}
\usepackage[colorlinks=true,citecolor=Blue,urlcolor=Blue,linkcolor=Blue]{hyperref}
\usepackage{microtype}
\usepackage{placeins}
\usepackage{comment}

\graphicspath{{Figure/}}
\newcommand{\E}{\mathbb{E}}
\newcommand{\Normal}{\mathcal{N}}
\newcommand{\dd}{\mathrm{d}}
\newcommand{\avQ}[1]{\left\langle #1\right\rangle_Q}

\begin{document}

\title{Phase Transitions in Mismatched Bayesian Inference for the Ising Model}

\author{Xinyu Sun}
\affiliation{Chen-Ning Yang Institute for Advanced Study (YIAS), Tsinghua University, Beijing 100084, China}

\author{Zhou-Quan Wan}
\affiliation{Center for Computational Quantum Physics, Flatiron Institute, New York, New York 10010, USA}

\author{Shao-Kai Jian}
\email{sjian@tulane.edu}
\affiliation{Department of Physics and Engineering Physics, Tulane University, New Orleans, Louisiana 70118, USA}

\author{Hong Yao}
\email{yaohong@tsinghua.edu.cn}
\affiliation{Chen-Ning Yang Institute for Advanced Study (YIAS), Tsinghua University, Beijing 100084, China}

\date{\today}

\begin{abstract}
Bayesian inference with a mismatched model can develop an overconfident phase, characterized by posterior agreement without faithful reconstruction of the true state. 
We study Bayesian inference from noisy bond measurements of an Ising model in its paramagnetic phase, with true inverse temperature $\beta$ and measurement-noise strength $\Delta$, while inference assumes corresponding parameters $\beta_Q$ and $\Delta_Q$. 
Using replica field theory, we analytically show that the overconfident phase can occur when inference underestimates the measurement noise ($\Delta_Q < \Delta$) or assumes a lower prior temperature than the true one ($\beta_Q > \beta$).
An exact change of variables maps the problem to a correlated random-bond Ising model, where the informative, uninformative, and overconfident phases correspond to ferromagnetic, paramagnetic, and spin-glass phases, respectively.
We further perform Monte Carlo simulations of the mapped model in three dimensions and establish a finite overconfident region between the informative and uninformative phases, consistent with our analytical prediction. 
Finally, we show that the true model parameters can be inferred directly from the measurement records, providing a natural route to calibrating the inference model.
\end{abstract}

\maketitle

{\it Introduction.---}Bayesian inference of interacting degrees of freedom admits a natural statistical-mechanics formulation: the prior and posterior are Gibbs ensembles, and changes in reconstructability can appear as thermodynamic phase transitions of the corresponding effective model~\cite{jaynes1957information,jaynes1957information2,nishimori2001book,iba1999nishimori,zdeborova2016statistical,mezard2009information,watkin1993statistical,seung1992statistical,opper1991generalization}. 
This viewpoint connects planted spin glasses, coding, and high-dimensional statistical inference~\cite{edwards1975theory,sherrington1975solvable,krzakala2012statistical,lesieur2017constrained,barbier2019mutual,gallager1962ldpc,mackay2003information,sourlas1989spin,sourlas1994finite,nishimori1999image,RicciTersenghi2019Typology}, and has become central to learning and information transitions induced by local measurements in classical and quantum many-body systems~\cite{Abbe2018Group,nahum2025bayesian,wiese2026bayesian,putz2026learning,patil2026learning,mcculloch2026bayesian,fisher2023random,jian2020measurement,bao2020theory,gullans2020scalable,garratt2023measurements,lee2023quantum,murciano2023measurement,barratt2022transitions,agrawal2024observing,ippoliti2024learnability,singh2026mixed,choi2020quantum,zhu2023nishimori,fan2024diagnostics,chen2025nishimori,wan2026revisiting}. 
Related statistical-mechanical structures also underlie fault-tolerant quantum memories~\cite{dennis2002topological,kitaev2003fault,Wang2003Confinement}.
Interestingly, Ref.~\cite{nahum2025bayesian} developed a finite-dimensional replica Landau framework for Bayes-optimal inference and emphasized an enlarged replica symmetry between the true configuration and posterior replicas, a structure closely tied to Nishimori identities~\cite{nishimori1981internal,nishimori2001book}.

Real inference, however, is rarely perfectly matched.
The prior temperature and measurement noise must be inferred from data. 
Once the assumed parameters differ from their true values, different notions of successful learning need no longer coincide. 
Independent posterior samples may develop macroscopic agreement even when they fail to faithfully reconstruct the true configuration.
The posterior can therefore develop internal agreement before becoming informative.
Related forms of model mismatch and nontrivial posterior ordering have appeared in inference and decoding problems~\cite{chen2022planted,camilli2022an,temkin2025charge,kim2026measurement,yan2026nonlinear,dellabate2026bayesian}.
However, whether parameter mismatch can split a single learning transition into distinct reconstruction and posterior-agreement transitions, thereby stabilizing a phase with posterior order but no faithful reconstruction, remains unclear.

Here we answer this question in the affirmative. 
In a replicated description, reconstruction and posterior agreement correspond to distinct truth--posterior and posterior--posterior overlap fields, which are symmetry-related under Bayes matching but respond differently to model mismatch. 
We use replica field theory to show that underestimating the measurement noise or assuming a lower prior temperature preferentially drives the posterior--posterior channel toward ordering, providing a mechanism for the overconfident phase.
The universality classes between different phases are also discussed and presented in Supplemental Material (SM)~\cite{SM}. 
We show that an exact change of variables further maps the problem to a correlated random-bond Ising model, where reconstruction and posterior agreement acquire the familiar interpretation of ferromagnetic and glassy overlap order. 
Our Monte Carlo simulations in three dimensions confirm that the two transitions split and bound a finite overconfident regime for both noise and temperature mismatch, in agreement with the analytical prediction. 
Finally, we show that the true parameters can be learned from the measurement records, either directly from low-order moments in the Gaussian model or more generally through evidence maximization.

\begin{figure}[t]
    \centering
    \includegraphics[width=\columnwidth]{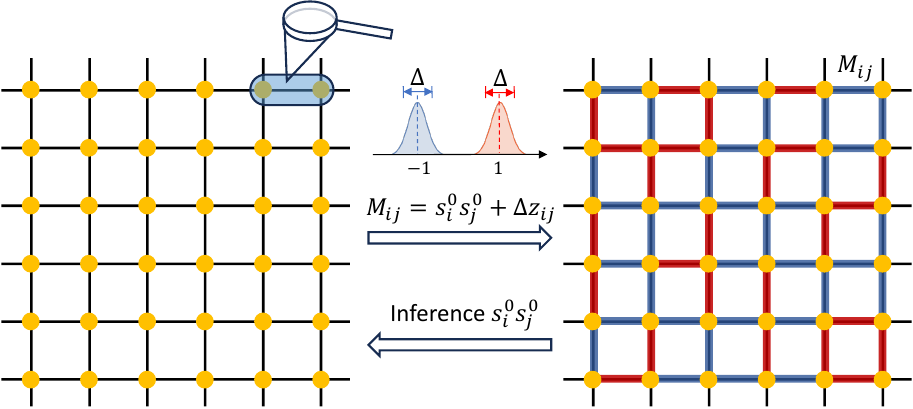}
    \caption{
    A true configuration $\bm s^0$ generates the record $M_{ij}=s_i^0s_j^0+\Delta z_{ij}$. 
    Inference assigns a posterior probability to candidate configurations conditioned on the same record.}
\label{fig:learning}
\end{figure}

{\it Model and order parameters.---}Let $s_i^0=\pm1$ denote the true Ising configuration on a $d$-dimensional lattice with nearest-neighbor bond set $E$.
It is drawn at inverse temperature $\beta$ according to
\begin{equation}
p_\beta(\bm s^0)=\frac{e^{\beta\sum_{\langle ij\rangle\in E}s_i^0s_j^0}}{Z_\beta}.
\label{eq:true_prior}
\end{equation}
Each bond variable is measured with independent Gaussian noise, yielding $M_{ij}=s_i^0s_j^0+\Delta z_{ij}$, where $z_{ij}\sim\Normal(0,1)$ and $\Delta$ is the true noise strength. 
The resulting joint distribution of the true configuration and measurement record is $p_{\beta,\Delta}(\bm s^0,\bm M)=p_\beta(\bm s^0)\prod_{\langle ij\rangle\in E}p_\Delta(M_{ij}|s_i^0s_j^0)$, where $p_\Delta(M_{ij}|s_i^0s_j^0)$ is a Gaussian distribution with mean $s_i^0s_j^0$ and variance $\Delta^2$.

Given a measurement record $\bm M$, inference assigns a posterior probability to each candidate configuration $\bm s$ using an assumed inverse temperature $\beta_Q$ and noise strength $\Delta_Q$.
The resulting posterior distribution is
\begin{equation}
    Q(\bm s|\bm M)=\frac{1}{Z_Q(\bm M)}\exp\!\left[\sum_{\langle ij\rangle}\left(\beta_Q+\frac{M_{ij}}{\Delta_Q^2}\right)s_is_j\right],
\label{eq:posterior}
\end{equation}
where $Z_Q(\bm M)$ normalizes the distribution over $\bm s$. 
Equation~\eqref{eq:posterior} follows from Bayes' rule for the assumed Ising prior and Gaussian noise model, with terms independent of $\bm s$ absorbed into the normalization; see the SM for details~\cite{SM}.
When the assumed parameters match the true ones, $\beta_Q=\beta,\ \Delta_Q=\Delta$, the posterior is Bayes optimal.

For a fixed measurement record $\bm M$, we denote by $\langle\cdot\rangle_Q$ the expectation with respect to the posterior in Eq.~\eqref{eq:posterior}, and by $\E$ the average over the true joint distribution $p_{\beta,\Delta}(\bm s^0,\bm M)$. 
The two central correlations are
\begin{equation}
    C_P(i,j)=\E\!\left[s_i^0s_j^0\avQ{s_is_j}\right],\ \
    C_Q(i,j)=\E\!\left[\avQ{s_is_j}^{\,2}\right].
\label{eq:orders}
\end{equation}
Here $C_P$ measures truth--posterior reconstruction, namely whether the inferred relative orientation agrees with that of the true configuration, while $C_Q$ measures posterior--posterior agreement.  
We denote the long-distance limits of $C_P$ and $C_Q$ by $q_P$ and $q_Q$, respectively. 
We call a phase with $q_P,q_Q>0$ \emph{informative}, one with $q_P=q_Q=0$ \emph{uninformative}, and one with $q_P=0$ and $q_Q>0$ \emph{overconfident}. 
In the overconfident phase, posterior configurations exhibit mutual agreement without carrying information about the true configuration.

Bayes optimality imposes an exact constraint on these phases. 
Let $\langle\cdot\rangle_{\rm B}$ denote the posterior expectation at the true parameters $(\beta,\Delta)$, and $\E_M$ the average over the marginal distribution of measurement records. 
Since $\langle s_is_j\rangle_{\rm B}$ is the conditional expectation of the true bond $s_i^0s_j^0$ at fixed $\bm M$, $C_P(i,j)=\E_M[\langle s_is_j\rangle_{\rm B}\avQ{s_is_j}]$.
Defining $C_{\rm B}(i,j)=\E_M[\langle s_is_j\rangle_{\rm B}^{\,2}]$, the Cauchy--Schwarz inequality, together with $C_Q(i,j)\le1$, gives (see SM~\cite{SM})
\begin{equation}
    |C_P(i,j)|^2\le C_{\rm B}(i,j)C_Q(i,j)\le C_{\rm B}(i,j).
\label{eq:bayes_bound}
\end{equation}
Thus mismatch cannot produce long-range reconstruction where Bayes-optimal inference is uninformative, so the informative region under mismatch is contained in the Bayes-informative region.
Moreover, Eq.~\eqref{eq:bayes_bound} implies that $q_Q=0$ necessarily gives $q_P=0$, excluding a phase with $q_P>0$ but $q_Q=0$,  which means if the reconstruction succeeds, the posterior replicas must reach agreement.

{\it Replica field theory.---}To analyze the two ordering channels under mismatch, we use a replica formulation since the normalization $Z_Q(\bm M)$ fluctuates with the record.  
Let \(w_Q(\bm s;\bm M)\) denote the unnormalized weight in Eq.~\eqref{eq:posterior}. 
For an observable $\mathcal O$ that depends on the true configuration and $k$ conditionally independent posterior samples,
\begin{align}
    &\E\left\langle\mathcal O\right\rangle_{Q^{\otimes k}}
    =\lim_{n\to0}\sum_{\bm s^0,\bm s^1,\ldots,\bm s^n}
    \int\!\dd\bm M\,p_{\beta,\Delta}(\bm s^0,\bm M)\nonumber\\[-2pt]
    &\hspace{17mm}\times
    \mathcal O(\bm s^0,\bm s^1,\ldots,\bm s^k)
    \prod_{\alpha=1}^{n}w_Q(\bm s^\alpha;\bm M).
\label{eq:replica_identity}
\end{align}
For integer $n\ge k$, the $n-k$ spectator replicas generate $Z_Q^{n-k}$, and continuation to $n\to0$ restores the required factor $Z_Q^{-k}$. 
We keep the true configuration as replica $0$, while $\alpha=1,\ldots,n$ labels posterior replicas. 
Hence, $C_P$ is represented by $\mathcal O_P=s_i^0s_j^0s_i^\alpha s_j^\alpha$, while $C_Q$ is represented by $\mathcal O_Q=s_i^\alpha s_j^\alpha s_i^\gamma s_j^\gamma$ with $\alpha\ne\gamma$.

Averaging over the Gaussian measurement record in Eq.~\eqref{eq:replica_identity} gives the replicated lattice Hamiltonian (see SM~\cite{SM})
\begin{align}
    -\mathcal H_n={}&\beta\sum_{\langle ij\rangle}s_i^0s_j^0
    +\beta_Q\sum_{\alpha,\langle ij\rangle}s_i^\alpha s_j^\alpha +\lambda_P\sum_{\alpha,\langle ij\rangle}(s_i^0s_i^\alpha)(s_j^0s_j^\alpha)\nonumber\\
    &+\lambda_Q\sum_{\alpha<\gamma,\langle ij\rangle}(s_i^\alpha s_i^\gamma)(s_j^\alpha s_j^\gamma),
\label{eq:replica_hamiltonian}
\end{align}
where $\lambda_P=\Delta_Q^{-2}$ and $\lambda_Q=\Delta^2\Delta_Q^{-4}$ are inter-replica couplings. 
At matching, $\beta_Q=\beta$ and $\lambda_P=\lambda_Q$, restoring an enlarged $S_{n+1}$ symmetry among the true and posterior replicas. 

To obtain an effective theory for the two overlap channels, we apply a Hubbard--Stratonovich transformation and introduce overlap fields $\Psi_{P,i}^\alpha$ and $\Psi_{Q,i}^{\alpha\gamma}$ conjugate to $s_i^0s_i^\alpha$ and $s_i^\alpha s_i^\gamma$, respectively. 
Integrating out the Ising spins then generates an effective action for these overlap fields. 
To leading orders near the upper critical dimension $d_c=6$, where the cubic interactions are marginal and higher-order terms are irrelevant, the effective action takes the form,
\begin{align}
    S_{\rm eff}={}&\frac12\int_{\bm k}\left[\sum_\alpha\Gamma_P^{(2)}|\Psi_P^\alpha|^2+\sum_{\alpha<\gamma}\Gamma_Q^{(2)}|\Psi_Q^{\alpha\gamma}|^2\right]\label{eq:exact_RPA_kernel}\\
    &-\int_{\bm r}\left[w_1\sum_{\alpha<\gamma}\Psi_P^\alpha\Psi_Q^{\alpha\gamma}\Psi_P^\gamma+w_2\sum_{\alpha<\gamma<\delta}\Psi_Q^{\alpha\gamma}\Psi_Q^{\gamma\delta}\Psi_Q^{\alpha\delta}\right].\nonumber
\end{align}
where $\int_{\bm r}\equiv \int d^d r $, $\int_{\bm k} \equiv \int \frac{d^dk}{(2\pi)^d}$ and the coefficients $w_1$ and $w_2$ describe the leading cubic interactions. 
$\Gamma_X^{(2)}(\bm k)=\Lambda_X(\bm k)^{-1}-\chi_X^{(0)}(\bm k)$ for $X=P,Q$, where the first term is the bare Hubbard--Stratonovich kernel inherited from the inter-replica coupling, and $\chi_X^{(0)}$ is the corresponding susceptibility generated by the decoupled Ising sectors.
We have $\Lambda_X(\bm k)=\lambda_X\gamma(\bm k)$ with $\gamma(\bm k)=2\sum_{\ell=1}^d\cos k_\ell$, and 
\begin{equation}
\begin{split}
    \chi_P^{(0)}(\bm k)&=\sum_{\bm r}e^{i\bm k\cdot\bm r}C_\beta(\bm r)C_{\beta_Q}(\bm r),\\
    \chi_Q^{(0)}(\bm k)&=\sum_{\bm r}e^{i\bm k\cdot\bm r}C_{\beta_Q}(\bm r)^2,
\end{split}
\label{eq:lattice_overlap_chi}
\end{equation}
where $C_{\beta'}(\bm r)=\langle s_{\bm0}s_{\bm r}\rangle_{\beta'}$ is the clean-Ising correlator at inverse temperature $\beta'$ in the paramagnetic phase. 

The condition $\Gamma_X^{(2)}(0)=0$ marks the onset of ordering in the corresponding overlap channel. 
Away from Bayes matching, the replicated theory has symmetry $G=\mathbb Z_2^{(0)}\times\big[(\mathbb Z_2)^n\rtimes S_n\big]$, where $\mathbb Z_2^{(0)}$ flips the true replica, $s_i^0\to \epsilon_0s_i^0$, the $n$ factors in $(\mathbb Z_2)^n$ independently flip the posterior replicas, $s_i^\alpha\to \epsilon_\alpha s_i^\alpha$, and $S_n$ permutes the posterior replica labels. Accordingly, the overlap fields transform as
$\Psi_P^\alpha\to \epsilon_0\epsilon_\alpha\Psi_P^\alpha$, $\Psi_Q^{\alpha\gamma}\to \epsilon_\alpha\epsilon_\gamma\Psi_Q^{\alpha\gamma}$.
The uninformative, overconfident, and informative phases are characterized by $(\langle\Psi_P\rangle,\langle\Psi_Q\rangle)=(0,0)$, $(0,\neq0)$, and $(\neq0,\neq0)$, respectively, and can equivalently be distinguished by their symmetry-breaking patterns. 
The uninformative phase preserves $G$. 
Condensation of $\Psi_Q$ locks the independent posterior $\mathbb Z_2$ transformations to a common flip, leaving $ G_{\rm over}=\mathbb Z_2^{(0)}\times\mathbb Z_2^{(Q)}\times S_n$, while condensation of $\Psi_P$ further locks the true replica to the posterior replicas, leaving $G_{\rm info}=\mathbb Z_2^{\rm diag}\times S_n$. 
$\Gamma_P^{(2)}(0)$ and $\Gamma_Q^{(2)}(0)$ need not vanish simultaneously, allowing the two ordering transitions to split. 
The cubic coupling $w_1\Psi_P\Psi_Q\Psi_P$, however, precludes a phase with reconstruction but no posterior agreement.
Once $\Psi_P$ condenses, it generates a term linear in $\Psi_Q$, thereby inducing $\langle\Psi_Q\rangle\neq0$, consistent with the nonperturbative constraint in Eq.~\eqref{eq:bayes_bound}.
The effective critical theories and universality classes associated with these symmetry-breaking transitions are discussed in detail in the SM~\cite{SM}.

It becomes particularly transparent along the two mismatch cuts below.
For noise mismatch with $\beta_Q=\beta$, the bare susceptibilities are identical, $\chi_P^{(0)}=\chi_Q^{(0)}$, while $\Lambda_Q/\Lambda_P=\Delta^2/\Delta_Q^2$.
Hence, underestimating the noise, $\Delta_Q<\Delta$, directly enhances the $\Psi_Q$ channel relatively and favors posterior agreement over reconstruction.
Note that the same framework extends to a critical underlying Ising model, where the critical Ising fields $\phi^a$ must be retained, as briefly discussed in the SM~\cite{SM}.

For temperature mismatch with $\Delta_Q=\Delta$, the overlap couplings are identical, and the splitting instead arises from clean-Ising correlations.
For $\beta_Q>\beta$, ferromagnetic correlation inequalities imply $C_{\beta_Q}(\bm r)\ge C_\beta(\bm r)\ge0$, and $\chi_Q^{(0)}(0)\ge\chi_P^{(0)}(0)$. 
Thus a lower assumed prior temperature preferentially enhances the $\Psi_Q$ channel.
On a $d$-dimensional hypercubic lattice, 
the difference between the Gaussian instability thresholds satisfies
\begin{equation}
    \Delta_{c,Q}^2-\Delta_{c,P}^2=2d\left[\chi_Q^{(0)}(0)-\chi_P^{(0)}(0)\right].
\label{eq:rpa_transition_split}
\end{equation}
Writing $\delta\beta=\beta_Q-\beta$ and denoting by $\bm r_{\rm NN}$ a nearest-neighbor displacement vector, the high-temperature expansion $C_\beta(\bm r_{\rm NN})=\beta+O(\beta^3)$ gives  $\chi_Q^{(0)}(0)-\chi_P^{(0)}(0)=2d[\beta\,\delta\beta+(\delta\beta)^2]+\cdots$.
The splitting therefore grows linearly with $\delta\beta$ at small finite $\beta$, while at $\beta=0$ its leading dependence is quadratic in $\beta_Q$.

The cubic coupling has upper critical dimension $d_c=6$, above which the analytical treatment is controlled. 
However, we expect our qualitative result to persist in $d=3$, the lowest integer dimension allowing a finite-temperature spin-glass phase, and test this expectation numerically in the three dimensional Ising model below.
The result of mismatched Bayes inference for two dimensional Ising model is included in the SM~\cite{SM}, where no overconfident phase is found.

\begin{figure}[t]
    \centering
    \includegraphics[width=\columnwidth]{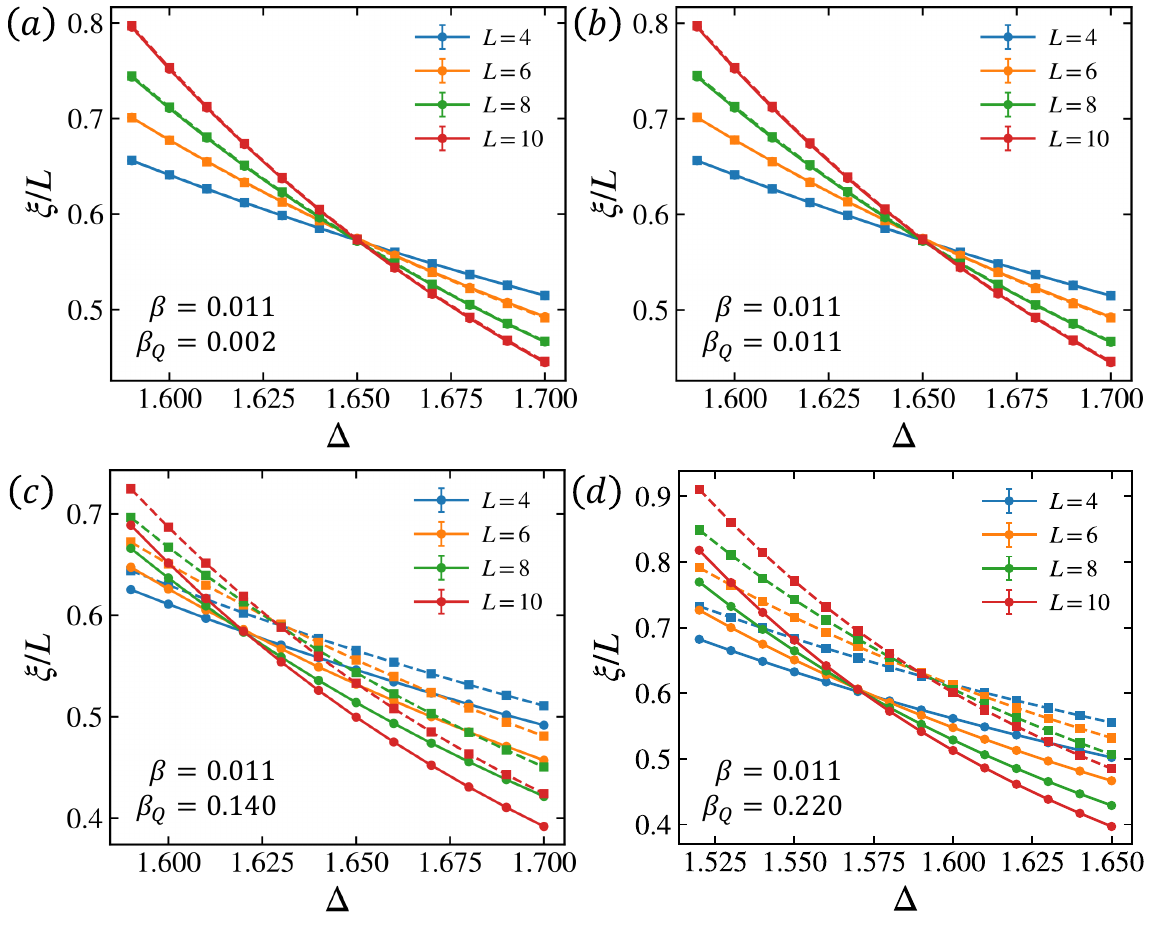}
    \caption{Finite-size crossings under temperature mismatch.
    We fix $\beta=0.011$ and matched noise $\Delta=\Delta_Q$, and vary $\Delta$ for (a) $\beta_Q=0.002$, (b) $\beta_Q=0.011$, (c) $\beta_Q=0.14$, and (d) $\beta_Q=0.22$. 
    Solid curves show $\xi_P/L$, associated with FM order (truth--posterior reconstruction), while dashed curves show $\xi_Q/L$, associated with SG order (posterior--posterior overlap). 
    Different colors denote $L=4,6,8,10$, with $N=35840$ disorder realizations for each data point.}
\label{fig:crossings}
\end{figure}

\begin{figure}[t]
\centering
    \includegraphics[width=\linewidth]{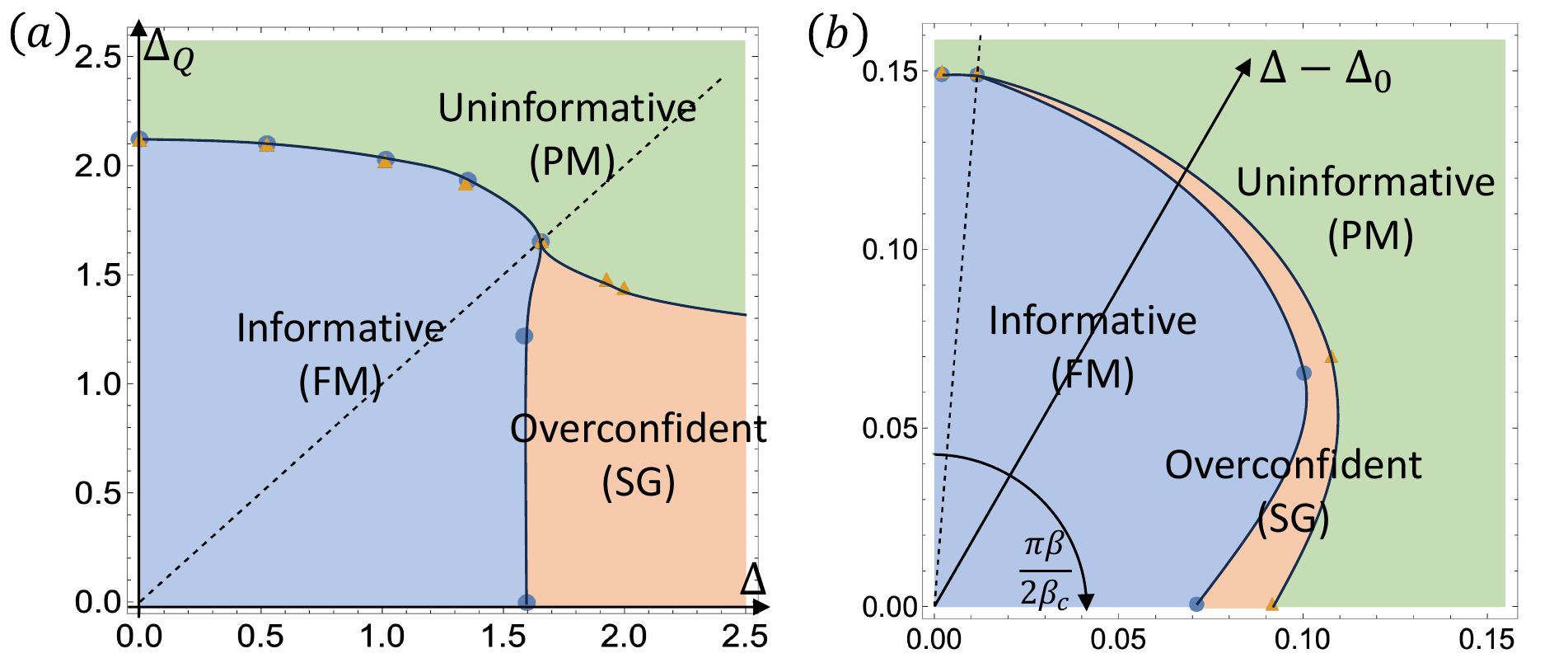}
    \caption{Numerical phase diagrams of mismatched Ising inference in 3D.
    (a) Noise mismatch at $\beta=\beta_Q=0$.
    We fix $\Delta/\Delta_Q=0.25,0.5,0.7,1,1.3$ and vary $\Delta_Q$ to determine the FM and SG transition boundaries. 
    In addition, we fix $\Delta=2$ and vary $\Delta_Q$ to locate the SG--PM transition. 
    Blue circles mark transitions detected by FM order, while yellow triangles mark those detected by SG order.
    The dashed line $\Delta=\Delta_Q$ is the Gaussian Nishimori line.
    The colored regions denote the informative (FM), overconfident (SG), and uninformative (PM) phases.
    (b) Temperature mismatch at fixed $\beta=0.011$ and matched noise $\Delta=\Delta_Q$, for $\beta_Q=0.002,0.011,0.14,0.22$. 
    To display the small splitting more clearly, we use polar-like coordinates with radius $\Delta-\Delta_0$, where $\Delta_0=1.5$, and angle $\theta=\pi\beta_Q/(2\beta_c)$, with $\beta_c=0.22165$.
    Thus the vertical and horizontal axes correspond to $\beta_Q=0$ and $\beta_Q=\beta_c$, respectively.
    The dashed ray marks the Nishimori condition $\beta_Q=\beta$.
    Marker colors have the same meaning as in panel (a). 
    The solid curves interpolate the numerical phase boundaries.}
\label{fig:phase}
\end{figure}

{\it Three-dimensional Ising inference.---}Consider a three-dimensional simple-cubic lattice, and define relative spin $\sigma_i=s_i^0s_i$.
Since multiplying the Gaussian noise by the fixed sign $s_i^0s_j^0$ leaves its distribution unchanged, Eq.~\eqref{eq:posterior} becomes a random bond Ising model (RBIM) for the relative spins,
\begin{equation}
    Q(\bm\sigma|\bm s^0,\bm z)\propto e^{\sum_{\langle ij\rangle}J_{ij}\sigma_i\sigma_j},
    \ J_{ij}=\frac1{\Delta_Q^2}+\beta_Qs_i^0s_j^0+\frac{\Delta}{\Delta_Q^2}z_{ij}.
\label{eq:gauge}
\end{equation}
The correlations in Eq.~\eqref{eq:orders} become $C_P(i,j)=\E[\langle\sigma_i\sigma_j\rangle_Q]$ and $C_Q(i,j)=\E[\langle\sigma_i\sigma_j\rangle_Q^2]$.
Thus $q_P$ corresponds to ferromagnetic (FM) long-range order, while $q_Q$ corresponds to two-replica overlap order. 
In particular, the overconfident regime $q_P=0$, $ q_Q>0$, with replica-overlap order but no FM order, is a spin-glass (SG) phase.

This correspondence is especially transparent in the limit $\beta=\beta_Q=0$, where Eq.~\eqref{eq:gauge} reduces exactly to an independent and identically distributed (iid) Gaussian RBIM with $\overline J=\Delta_Q^{-2}$ and ${\rm Var}(J)=\Delta^2\Delta_Q^{-4}$.
Note that ${\rm Var}(J)/\overline J=(\Delta/\Delta_Q)^2$. 
The Gaussian Nishimori line is therefore $\Delta=\Delta_Q$.
This provides a microscopic interpretation of the field-theory result above. 
The condition $\Delta>\Delta_Q$ corresponds to increasing bond randomness relative to the ferromagnetic bias and moving from the Nishimori line toward the SG side.
The SG order can therefore persist after FM order is lost, giving the overconfident phase.

We define the observables in the simulation for the relative-spin RBIM. 
We denote by $\xi_P$ the correlation length of the relative spins and by $\xi_Q$ that of $q_i=\sigma_i^1\sigma_i^2$, where $\bm\sigma^1$ and $\bm\sigma^2$ are two independent replicas with the same disorder realization.
From the corresponding structure factors $\chi_P$ and $\chi_Q$,
\begin{equation}
    \xi_X=\frac{1}{2\sin(\pi/L)}\left[\frac{\chi_X(0)}{\chi_X(\bm k_{\min})}-1\right]^{1/2},
    \quad X=P,Q,
\label{eq:xi}
\end{equation}
where $\bm k_{\min}=(2\pi/L,0,0)$. 
Crossings of $\xi_P/L$ and $\xi_Q/L$ locate the FM and SG transitions, respectively. 
The structure factors and numerical estimators are detailed in the SM~\cite{SM}.

We first examine the temperature-mismatch cut at fixed $\beta=0.011$ with $\Delta=\Delta_Q$.
Figure~\ref{fig:crossings} shows the finite-size crossings of $\xi_P/L$ and $\xi_Q/L$ for different $\beta_Q$.
For $\beta_Q\le\beta$, the FM and SG crossing families coincide within the available sizes. 
Once $\beta_Q>\beta$, the two crossings separate clearly, with the FM transition occurring before the SG transition as the noise is increased.
Therefore, the intermediate SG interval provides direct numerical evidence for the overconfident phase. 
This ordering is consistent with the analytical prediction that a lower assumed prior temperature preferentially enhances posterior agreement.

For noise mismatch at $\beta=\beta_Q=0$, the numerical results also show the predicted  behavior. 
The FM and SG transitions remain coincident for $\Delta\le\Delta_Q$, while for $\Delta>\Delta_Q$ they separate and produce an intermediate overconfident phase. 
The corresponding finite-size crossing data are given in the SM~\cite{SM}.

Collecting these crossing estimates gives the numerical phase diagrams in Fig.~\ref{fig:phase}. 
At $\beta=\beta_Q=0$, Fig.~\ref{fig:phase}(a) exhibits the informative (FM), overconfident (SG), and uninformative (PM) phases~\footnote{The transition point on the $\Delta_Q=0$ axis, corresponding to the zero-temperature three-dimensional Gaussian RBIM, is taken from Ref.~\cite{Krzakala2002Absence}. The transition point on the $\Delta=0$ axis is determined by the critical coupling of the pure three-dimensional Ising model~\cite{Ferrenberg2018Pushing}.}.
The overconfident region lies on the $\Delta>\Delta_Q$ side of the Nishimori line, as shown in Fig.~\ref{fig:phase}(a). 
In addition, Figure~\ref{fig:phase}(b) shows the temperature-mismatch phase diagram at matched noise, where the overconfident region appears for $\beta_Q>\beta$. 
Hence, both the phase structure and the location of the overconfident region agree with the analytical prediction.

{\it Learning the unknown parameters.---}Having characterized the consequences of mismatch, we finally ask whether the true parameters can themselves be inferred from the measurement records. 
For the present Gaussian channel, this can be done directly. 
Define the bond-averaged record moments $m_k=N_b^{-1}\sum_{\langle ij\rangle}\E[M_{ij}^k]$, where $N_b$ is the number of bonds, and the clean-Ising bond energy density $e(\beta)=N_b^{-1}\langle\sum_{\langle ij\rangle}s_is_j\rangle_\beta$. 
Since $M_{ij}=s_i^0s_j^0+\Delta z_{ij}$ with $z_{ij}\sim{\cal N}(0,1)$, the first two moments satisfy
\begin{equation}
    m_1=e(\beta),\qquad m_2=1+\Delta^2.
\label{eq:parameter_moments}
\end{equation}
Thus $\Delta^2=m_2-1$ determines the noise strength independently of $\beta$. 
The inverse temperature follows from $m_1$, because $\partial_\beta e(\beta)=N_b^{-1}{\rm Var}_\beta(\sum_{\langle ij\rangle}s_is_j)\ge0$ and $e(\beta)$ is monotonic.
Once the clean equation of state is known, the first two record moments determine $\beta$ and $\Delta$ separately, allowing the true parameters to be recovered without posterior inference.

More generally, when such moment identities are unavailable, the parameters can be estimated by evidence maximization. 
Let $p_Q(\bm M)$ denote the probability assigned to a measurement record $\bm M$ by the assumed model with parameters $(\beta_Q,\Delta_Q)$. 
For $R$ independent records $\{\bm M^{(r)}\}_{r=1}^R$, where $r$ labels the record, we determine the parameters by maximizing the average log evidence, $\frac1R \sum_{r=1}^R \ln p_Q({\bm M}^{(r)})$. 
In the large-$R$ limit, this procedure recovers the true parameters $(\beta,\Delta)$. 
Equivalently, evidence maximization minimizes the Kullback--Leibler divergence between the true and assumed record distributions; see the SM~\cite{SM} for the derivation and the corresponding self-consistency equations.

{\it Concluding remarks.---}We have established the mismatched Ising inference problem as a concrete and comprehensive setting in which the thermodynamics of confidence and reconstruction can be studied separately.
Our results provide an analytically and numerically tractable framework for understanding how confidence and information can decouple in mismatched inference. 
We further show in the SM that the same mismatched-inference structure extends to the deformed toric code, where logical learning maps onto ordering in a dual RBIM and no stable finite-temperature overconfident phase is expected in two dimensions.

Several directions remain open.
In particular, it will be interesting to investigate numerically the critical behavior of the correlated RBIM generated by temperature mismatch, including how the disorder correlations modify the split transitions and the overconfident phase.
More broadly, a central question is whether the separation between confidence and reconstruction persists beyond the present setting, and represents a generic phase structure of mismatched inference. 
It will also be important to understand how finite-data parameter estimation and residual model misspecification affect the location and observability of these transitions~\cite{Kleijn2006Misspecification,Nott2024Bayesian}.

{\it Acknowledgments:} We thank Adam Nahum for insightful discussions and valuable comments on an earlier version of the manuscript.
This work is supported in part by the NSFC under Grant Nos.~12347107 and 12334003 (X.S. and H.Y.), the New Cornerstone Science Foundation through the Xplorer Prize (H.Y.). 
The Flatiron Institute is a division of the Simons Foundation.
We used the AI tool GPT-5.6 to assist with figure generation and manuscript editing.

\let\oldaddcontentsline\addcontentsline
\renewcommand{\addcontentsline}[3]{}
\bibliography{prl_reference_v1}
\let\addcontentsline\oldaddcontentsline

\clearpage

\setcounter{section}{0}
\setcounter{subsection}{0}
\setcounter{secnumdepth}{2}
\setcounter{tocdepth}{2}

\renewcommand{\thesection}{\Roman{section}}
\renewcommand{\thesubsection}{\Alph{subsection}}
\renewcommand{\theHsection}{SM.\arabic{section}}
\renewcommand{\theHsubsection}{SM.\arabic{section}.\arabic{subsection}}

\setcounter{table}{0}
\renewcommand{\thetable}{S\arabic{table}}
\renewcommand{\theHtable}{S.\arabic{table}}

\setcounter{figure}{0}
\renewcommand{\thefigure}{S\arabic{figure}}
\renewcommand{\theHfigure}{S.\arabic{figure}}

\setcounter{equation}{0}
\renewcommand{\theequation}{S\arabic{equation}}
\renewcommand{\theHequation}{S.\arabic{equation}}

\begin{widetext}

\begin{center}
{\large\bf Supplemental Material for ``Phase Transitions in Mismatched Bayesian Inference for the Ising Model''}
\end{center}


\tableofcontents

\section{Posterior convention and exact Bayes inequality}
\label{sec:s_posterior_bayes}

We begin by fixing the probabilistic conventions used throughout the SM. 
We first distinguish the normalized joint distribution, the conditional posterior, and the marginal probability of the measurement record.
We then derive an exact Bayes inequality relating truth--posterior reconstruction and posterior--posterior agreement, which directly restricts the mismatched phase diagram.

\subsection{Posterior distribution and record probability}
\label{sec:s_posterior}

Conditioned on the true configuration $\bm s_i^0$, the true Gaussian measurement channel is
\begin{equation}
    p_\Delta(M_{ij}|s_i^0s_j^0)=\frac{1}{\sqrt{2\pi\Delta^2}}\exp\!\left[-\frac{(M_{ij}-s_i^0s_j^0)^2}{2\Delta^2}\right],
\label{eq:s_channel}
\end{equation}
and the normalized true joint distribution is
\begin{equation}
    p_{\beta,\Delta}(\bm s^0,\bm M)=\frac{e^{\beta\sum_{\langle ij\rangle} s_i^0s_j^0}}{Z_\beta}\prod_{\langle ij\rangle} p_\Delta(M_{ij}|s_i^0s_j^0).
\label{eq:s_true_joint}
\end{equation}
Inference assumes an Ising prior at inverse temperature $\beta_Q$ and a Gaussian measurement with uncertainty $\Delta_Q$.
The corresponding normalized assumed joint distribution is
\begin{equation}
    p_Q(\bm s,\bm M)=\frac{e^{\beta_Q\sum_{\langle ij\rangle}s_is_j}}{Z_{\beta_Q}}\prod_{\langle ij\rangle} p_{\Delta_Q}(M_{ij}|s_is_j).
\label{eq:s_assumed_joint}
\end{equation}
Throughout the SM, the subscript $Q$ on $p_Q$ indicates that the assumed generative model uses the parameters $(\beta_Q,\Delta_Q)$.
For the true model we instead write $p_{\beta,\Delta}$ explicitly.
On the matched manifold, $(\beta_Q,\Delta_Q)=(\beta,\Delta)$, and the two joint distributions coincide, $p_Q=p_{\beta,\Delta}$.

Using $(s_is_j)^2=1$, so that $(M_{ij}-s_is_j)^2=M_{ij}^2+1-2M_{ij}s_is_j$, gives
\begin{equation}
    p_Q(\bm s,\bm M)=\frac{1}{Z_{\beta_Q}(2\pi\Delta_Q^2)^{N_b/2}}
    \exp\!\left[-\frac{1}{2\Delta_Q^2}\sum_{\langle ij\rangle}(M_{ij}^2+1)\right]
    \exp\!\left[\sum_{\langle ij\rangle}\left(\beta_Q+\frac{M_{ij}}{\Delta_Q^2}\right)s_is_j\right],
\label{eq:s_assumed_joint_expanded}
\end{equation}
where $N_b$ is the number of bonds, and the first two terms depend on the observed record $\bm{M}$ but not on the candidate configuration $\bm{s}$.
It is useful to write this as $p_Q(\bm s,\bm M)=F_Q(\bm M)w_Q(\bm s;\bm M)$, where
\begin{equation}
    F_Q(\bm M)=\frac{e^{-\sum_{\langle ij\rangle}(M_{ij}^2+1)/(2\Delta_Q^2)}}{Z_{\beta_Q}(2\pi\Delta_Q^2)^{N_b/2}},\qquad 
    w_Q(\bm s;\bm M)=\exp\!\left[\sum_{\langle ij\rangle}\left(\beta_Q+\frac{M_{ij}}{\Delta_Q^2}\right)s_i s_j\right].
\label{eq:s_Fw}
\end{equation}
We use a semicolon in $w_Q(\bm s;\bm M)$ to emphasize that it is an unnormalized weight depending parametrically on the record $\bm M$, rather than a conditional probability.
The posterior is obtained by conditioning the assumed joint distribution on that fixed record,
\begin{equation}
    Q(\bm s|\bm M)=\frac{p_Q(\bm s,\bm M)}{\sum_{\bm s'}p_Q(\bm s',\bm M)}=\frac{w_Q(\bm s;\bm M)}{Z_Q(\bm M)},\qquad 
    Z_Q(\bm M)=\sum_{\bm s}w_Q(\bm s;\bm M).
\label{eq:s_conditional}
\end{equation}
Thus $F_Q(\bm M)$ cancels between numerator and denominator, which explains directly why the Gaussian normalization, the clean-prior normalization $Z_{\beta_Q}$, and the record-only factor in Eq.~\eqref{eq:s_assumed_joint_expanded} do not appear in $Q(\bm s|\bm M)$ of the main text. 
By contrast, these factors must be retained when one asks for the probability assigned to the record itself.
Summing the normalized joint distribution over the latent spins gives $p_Q(\bm M)=\sum_{\bm s}p_Q(\bm s,\bm M)=F_Q(\bm M)Z_Q(\bm M)$.
Hence $Z_Q(\bm M)$ is the partition function that normalizes the posterior over $\bm s$ for a fixed record $\bm M$, whereas $p_Q(\bm M)=F_Q(\bm M)Z_Q(\bm M)$ is the normalized probability density assigned by the assumed generative model to the record itself. 
The latter is the evidence used for parameter learning in Sec.~\ref{sec:s_parameter_learning}.


\subsection{Exact Bayes inequality}
\label{sec:s_bayes}

With these conventions fixed, we now derive an exact constraint relating reconstruction and posterior agreement. 
For a fixed measurement record $\bm M$, we denote by $\langle\cdots\rangle_Q$ the expectation value with respect to the mismatched posterior $Q(\bm s|\bm M)$ defined in Sec.~\ref{sec:s_posterior}. 
Similarly, $\langle\cdots\rangle_{\rm B}$ denotes the posterior expectation obtained using the true parameters $(\beta,\Delta)$. Let
\begin{equation}
    p_{\beta,\Delta}(\bm M)=\sum_{\bm s^0}p_{\beta,\Delta}(\bm s^0,\bm M)
\label{eq:s_record_marginal}
\end{equation}
be the true marginal distribution of measurement records. 
The optimal Bayes posterior is
\begin{equation}
    B(\bm s|\bm M)=p_{\beta,\Delta}(\bm s|\bm M)=\frac{p_{\beta,\Delta}(\bm s,\bm M)}{p_{\beta,\Delta}(\bm M)}.
\label{eq:s_bayes_posterior}
\end{equation}
Its bond expectation at fixed $\bm M$ is $\langle s_is_j\rangle_{\rm B}=\sum_{\bm s}s_is_j B(\bm s|\bm M)$.
Starting from the definition of the truth--posterior correlation,
\begin{align}
    C_P(i,j)&=\sum_{\bm s^0}\int\dd\bm M\,p_{\beta,\Delta}(\bm s^0,\bm M)s_i^0s_j^0\langle s_is_j\rangle_Q\nonumber\\
    &=\int\dd\bm M\,p_{\beta,\Delta}(\bm M)\left[\sum_{\bm s^0}s_i^0s_j^0B(\bm s^0|\bm M)\right]\langle s_is_j\rangle_Q\nonumber\\
    &=\int\dd\bm M\,p_{\beta,\Delta}(\bm M)\langle s_is_j\rangle_{\rm B}\langle s_is_j\rangle_Q.
\label{eq:s_CP_bayes}
\end{align}
The posterior--posterior and optimal Bayes reconstruction correlators can be written as
\begin{equation}
    C_Q(i,j)=\int\dd\bm M\,p_{\beta,\Delta}(\bm M)\langle s_is_j\rangle_Q^2,\qquad
    C_{\rm B}(i,j)=\int\dd\bm M\,p_{\beta,\Delta}(\bm M)\langle s_is_j\rangle_{\rm B}^2.
\label{eq:s_CQ_CB}
\end{equation}
Applying the Cauchy--Schwarz inequality directly to Eq.~\eqref{eq:s_CP_bayes} gives
\begin{equation}
    |C_P(i,j)|^2\leq\left[\int\dd\bm M\,p_{\beta,\Delta}(\bm M)\langle s_is_j\rangle_{\rm B}^2\right]\left[\int\dd\bm M\,p_{\beta,\Delta}(\bm M)\langle s_is_j\rangle_Q^2\right]=C_{\rm B}(i,j)C_Q(i,j).
\label{eq:s_bayes_CS}
\end{equation}
Since $|\langle s_is_j\rangle_Q|\leq1$, one has $C_Q(i,j)\leq1$, and therefore
\begin{equation}
    |C_P(i,j)|^2\leq C_{\rm B}(i,j)C_Q(i,j)\leq C_{\rm B}(i,j),
\label{eq:s_bayes_bound}
\end{equation}
which is shown in the main text.

Equation~\eqref{eq:s_bayes_bound} gives a direct constraint on the mismatched phase diagram. 
For fixed true parameters $(\beta,\Delta)$, whether long-range reconstruction is possible is bounded by the optimal Bayes posterior at those same true parameters. 
If $C_{\rm B}(i,j)$ vanishes at long distance, then $C_P(i,j)$ must also vanish for \emph{any} choice of assumed parameters $(\beta_Q,\Delta_Q)$. 
Thus mismatch cannot generate an informative phase once the corresponding optimal Bayes problem is already uninformative.
Moreover, if $C_Q(i,j)$ vanishes at long distance, the first inequality forces $C_P(i,j)$ to vanish as well, excluding a phase with $q_P>0$ and $q_Q=0$, which means if the reconstruction succeeds, the posterior replica must reach agreement.

This bound is particularly useful for the two mismatch cuts studied in the main text. 
\begin{itemize}
    \item For noise mismatch, fix the true inverse temperature $\beta$ and let $\Delta_c(\beta)$ denote the reconstruction threshold of the matched problem, defined by $\beta_Q=\beta$ and $\Delta_Q=\Delta$.
    If the true noise satisfies
    \begin{equation}
        \Delta>\Delta_c(\beta),
    \label{eq:s_noise_bound}
    \end{equation}
    the optimal Bayes problem is already uninformative. 
    The bound in Eq.~\eqref{eq:s_bayes_bound} therefore implies $q_P=0$ for any choice of the assumed noise $\Delta_Q$. 
    Equivalently, in the $(\Delta,\Delta_Q)$ plane at fixed $\beta_Q=\beta$, the informative region is confined to $\Delta\leq\Delta_c(\beta)$, regardless of $\Delta_Q$.

    \item For temperature mismatch, fix the true inverse temperature $\beta$ and impose matched noise $\Delta_Q=\Delta$. 
    The matched point $\beta_Q=\beta$ loses reconstruction at the same threshold $\Delta_c(\beta)$. 
    Therefore, if
    \begin{equation}
        \Delta=\Delta_Q>\Delta_c(\beta),
    \label{eq:s_temperature_bound}
    \end{equation}
    the optimal Bayes problem is uninformative and $q_P=0$ for any assumed inverse temperature $\beta_Q$. 
    Hence varying $\beta_Q$, whether to a higher or lower assumed temperature, cannot move the FM boundary to $\Delta>\Delta_c(\beta)$.
\end{itemize}

At matching, $(\beta_Q,\Delta_Q)=(\beta,\Delta)$, the assumed posterior is the optimal Bayes posterior, $Q=B$.
Equations~\eqref{eq:s_CP_bayes} and \eqref{eq:s_CQ_CB} then give
\begin{equation}
    C_P(i,j)=C_Q(i,j)=C_{\rm B}(i,j),
\label{eq:s_matching_correlations}
\end{equation}
so the Bayes bound is saturated on the Nishimori manifold.


\section{Replicated lattice theory}
\label{sec:s_replica}

We derive a replicated lattice description in which the truth--posterior and posterior--posterior overlaps appear as distinct interacting degrees of freedom.


The record-dependent normalization $Z_Q(\bm M)$ prevents a direct disorder average of normalized posterior moments.
To represent its inverse powers, we introduce posterior replicas.
Consider an observable $\mathcal O$ involving the true configuration $\bm s^0$ and $k$ conditionally independent posterior configurations $\bm s^1,\ldots,\bm s^k$. 
At fixed $\bm M$, its posterior expectation is
\begin{equation}
    \left\langle\mathcal O\right\rangle_{Q^{\otimes k}}=\frac{1}{Z_Q(\bm M)^k}\sum_{\bm s^1,\ldots,\bm s^k}\mathcal O(\bm s^0,\bm s^1,\ldots,\bm s^k)\prod_{\alpha=1}^k w_Q(\bm s^\alpha;\bm M).
\label{eq:s_posterior_moment}
\end{equation}
Averaging also over the true configuration and measurement record gives
\begin{equation}
    \E\left\langle\mathcal O\right\rangle_{Q^{\otimes k}}
    =\sum_{\bm s^0}\int\dd\bm M\,p_{\beta,\Delta}(\bm s^0,\bm M)\frac{\sum_{\bm s^1,\ldots,\bm s^k}\mathcal O(\bm s^0,\bm s^1,\ldots,\bm s^k)\prod_{\alpha=1}^k w_Q(\bm s^\alpha;\bm M)}{Z_Q(\bm M)^k}.
\label{eq:s_posterior_moment_average}
\end{equation}
The difficulty is the record-dependent denominator $Z_Q(\bm M)^k$.
To remove it, introduce $n-k$ additional posterior replicas.
For integer $n\geq k$, summing over these spectator replicas gives
\begin{equation}
    \sum_{\bm s^1,\ldots,\bm s^n}\mathcal O(\bm s^0,\bm s^1,\ldots,\bm s^k)\prod_{\alpha=1}^n w_Q(\bm s^\alpha;\bm M)=Z_Q(\bm M)^{n-k}\sum_{\bm s^1,\ldots,\bm s^k}\mathcal O(\bm s^0,\bm s^1,\ldots,\bm s^k)\prod_{\alpha=1}^k w_Q(\bm s^\alpha;\bm M).
\label{eq:s_spectator_identity}
\end{equation}
Analytically continuing the number of posterior replicas to $n\to0$ gives $Z_Q(\bm M)^{n-k}\to Z_Q(\bm M)^{-k}$ and therefore reproduces Eq.~\eqref{eq:s_posterior_moment_average}.
We obtain the replica representation
\begin{align}
    \E\left\langle\mathcal O\right\rangle_{Q^{\otimes k}}
    =\lim_{n\to0}\sum_{\bm s^0,\bm s^1,\ldots,\bm s^n}\int\dd\bm M\,p_{\beta,\Delta}(\bm s^0,\bm M)\,
    \mathcal O(\bm s^0,\bm s^1,\ldots,\bm s^k)\prod_{\alpha=1}^n w_Q(\bm s^\alpha;\bm M).
\label{eq:s_replica_identity}
\end{align}
Replica $0$ denotes the physical true configuration and is never analytically continued. 
The indices $\alpha,\gamma,\delta=1,\ldots,n$ label posterior replicas, whose number is analytically continued to zero.

The replica construction therefore treats the truth--posterior and posterior--posterior correlations within a single replicated statistical-mechanical model. 
For the two overlap correlations studied in the main text, $C_P$ contains one posterior sample and therefore corresponds to $\mathcal O_P=s_i^0s_j^0s_i^1s_j^1$, while $C_Q$ contains two independent posterior samples and corresponds to $\mathcal O_Q=s_i^1s_j^1s_i^2s_j^2$.



Because the measurement noise is Gaussian, we can integrate out the measurement record. 
After averaging over all measurement records, the exact replicated lattice Hamiltonian becomes
\begin{equation}
    -\mathcal H_n=\beta\sum_{\langle ij\rangle}s_i^0s_j^0 +\beta_Q\sum_{\alpha,\langle ij\rangle}s_i^\alpha s_j^\alpha +\lambda_P\sum_{\alpha,\langle ij\rangle}p_i^\alpha p_j^\alpha +\lambda_Q\sum_{\alpha<\gamma,\langle ij\rangle}q_i^{\alpha\gamma}q_j^{\alpha\gamma},
\label{eq:s_replica_hamiltonian}
\end{equation}
where $\lambda_P=\frac1{\Delta_Q^2}$, $\lambda_Q=\frac{\Delta^2}{\Delta_Q^4}$ and we introduce the microscopic site-overlap fields
\begin{equation}
    p_i^\alpha=s_i^0s_i^\alpha,\qquad
    q_i^{\alpha\gamma}=s_i^\alpha s_i^\gamma,\qquad \alpha\ne\gamma.
\label{eq:s_site_overlaps}
\end{equation}
In the replicated lattice Hamiltonian, Eq.~\eqref{eq:s_replica_hamiltonian}, the first term describes the true Ising configuration at inverse temperature $\beta$, while the second describes the posterior replicas at the assumed inverse temperature $\beta_Q$.
The record average generates the last two interactions.
The coupling $\lambda_P$ acts directly on the true-posterior overlap $p^\alpha$, while $\lambda_Q$ acts on the posterior-posterior overlap $q^{\alpha\gamma}$.


On the matched manifold, $\beta_Q=\beta$ and $\Delta_Q=\Delta$, so $\lambda_P=\lambda_Q=\frac1{\Delta^2}$.
The true configuration and all posterior replicas then have identical single-replica couplings and identical pairwise inter-replica couplings. 
The replicated Hamiltonian is therefore invariant under permutations of all $n+1$ replicas, making the enlarged $S_{n+1}$ symmetry explicit.
Away from matching, the single-replica couplings and the overlap couplings distinguish the $P$ and $Q$ sectors. 

\section{Replica Field theory}
\label{sec:s_continuum}

\subsection{Hubbard--Stratonovich transformation and overlap susceptibilities}

Starting from the exact replicated lattice theory above, we derive the replica field theory and analyze which channel is preferentially driven toward ordering under mismatch. 
We separate Eq.~\eqref{eq:s_replica_hamiltonian} into the decoupled single-replica Hamiltonian
\begin{equation}
    -\mathcal H_0=\beta\sum_{\langle ij\rangle}s_i^0s_j^0+\beta_Q\sum_{\alpha,\langle ij\rangle}s_i^\alpha s_j^\alpha
\label{eq:s_lattice_H0}
\end{equation}
and the inter-replica overlap interactions.

Define the nearest-neighbor kernels
\begin{equation}
    \Lambda_{P,ij}=\lambda_P\mathcal A_{ij},\qquad
    \Lambda_{Q,ij}=\lambda_Q\mathcal A_{ij},
\label{eq:s_lattice_Lambda}
\end{equation}
where $\mathcal A_{ij}=1$ for nearest-neighbor sites and vanishes otherwise.
In momentum space,
\begin{equation}
    \Lambda_P(\bm k)=\lambda_P\gamma(\bm k),\qquad
    \Lambda_Q(\bm k)=\lambda_Q\gamma(\bm k),\qquad
    \gamma(\bm k)=2\sum_{\ell=1}^d\cos k_\ell.
\label{eq:s_lattice_Lambda_k}
\end{equation}
We introduce fields $\Psi_{P,i}^\alpha$ and $\Psi_{Q,i}^{\alpha\gamma}$ conjugate directly to the local spin products $s_i^0s_i^\alpha$ and $s_i^\alpha s_i^\gamma$, and implement a lattice Hubbard--Stratonovich transformation.
The replicated partition function can then be written schematically as
\begin{equation}
    Z_n\propto\int\mathcal D\Psi_P\mathcal D\Psi_Q\,e^{-S_{\rm bare}[\Psi]}\sum_{\{s\}}e^{-\mathcal H_0}
    \exp\!\Bigg[\sum_{i,\alpha}\Psi_{P,i}^\alpha s_i^0s_i^\alpha+ \sum_{i,\alpha<\gamma}\Psi_{Q,i}^{\alpha\gamma}s_i^\alpha s_i^\gamma\Bigg],
\label{eq:s_lattice_HS_partition}
\end{equation}
with the bare Hubbard--Stratonovich action
\begin{equation}
    S_{\rm bare}[\Psi]=\frac12\sum_{\alpha,ij}\Psi_{P,i}^\alpha(\Lambda_P^{-1})_{ij}\Psi_{P,j}^\alpha
    +\frac12\sum_{\alpha<\gamma,ij}\Psi_{Q,i}^{\alpha\gamma}(\Lambda_Q^{-1})_{ij}\Psi_{Q,j}^{\alpha\gamma}.
\label{eq:s_lattice_HS_bare}
\end{equation}
Writing $\langle\cdots\rangle_0$ for the expectation value with respect to the decoupled Hamiltonian $\mathcal H_0$, integration over the lattice spins is encoded exactly by
\begin{equation}
    S_{\rm eff}[\Psi]
    =S_{\rm bare}[\Psi]-\ln\left\langle\exp\!\left[\sum_{i,\alpha}\Psi_{P,i}^\alpha s_i^0s_i^\alpha+\sum_{i,\alpha<\gamma}\Psi_{Q,i}^{\alpha\gamma}s_i^\alpha s_i^\gamma\right]\right\rangle_0.
\label{eq:s_lattice_effective_action}
\end{equation}

\begin{figure}[t]
    \centering
    \includegraphics[width=0.8\linewidth]{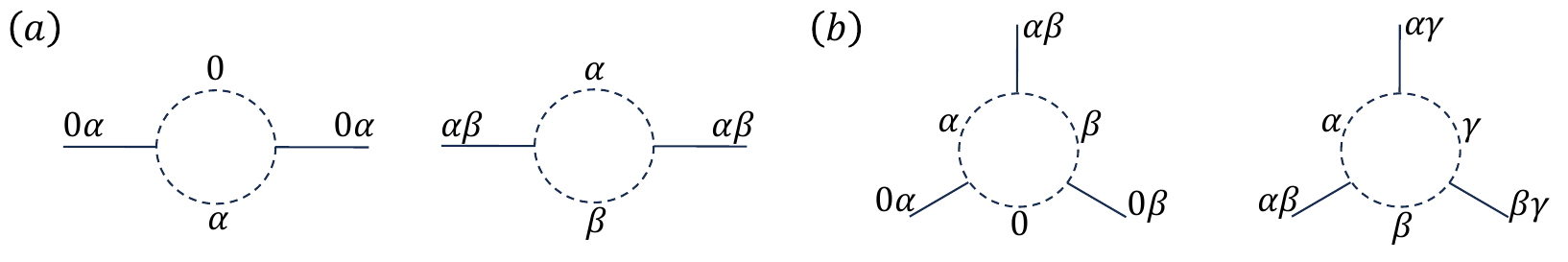}
    \caption{Feynman diagrams contributing to the quadratic and leading nonlinear terms of the replica field effective action.
    Solid lines denote the Hubbard--Stratonovich overlap fields $\Psi$, while dashed internal lines denote the underlying true and posterior correlation lines, given by exact clean-Ising spin correlations. 
    (a) Bubble diagrams for the truth--posterior field $\Psi_P^\alpha$ and posterior--posterior field $\Psi_Q^{\alpha\beta}$, which determine their quadratic kernels.
    (b) Three-point diagrams generating the cubic interactions $\Psi_P^\alpha\Psi_Q^{\alpha\beta}\Psi_P^\beta$ and $\Psi_Q^{\alpha\beta}\Psi_Q^{\beta\gamma}\Psi_Q^{\alpha\gamma}$.
    The labels on the dashed lines indicate the true replica $0$ and posterior replicas $\alpha,\beta,\gamma$.}
\label{fig:Feynman_diagram}
\end{figure}

We now expand Eq.~\eqref{eq:s_lattice_effective_action} to quadratic order in the Hubbard--Stratonovich fields. 
The linear term vanishes in the zero-field Ising theory. 
The two quadratic contractions, shown in Fig.~\ref{fig:Feynman_diagram}(a), are given by 
\begin{equation}
    \left\langle(s_{\bm0}^0s_{\bm0}^\alpha)(s_{\bm r}^0s_{\bm r}^\alpha)\right\rangle_0=C_\beta(\bm r)C_{\beta_Q}(\bm r),\qquad
    \left\langle(s_{\bm0}^\alpha s_{\bm0}^\gamma)(s_{\bm r}^\alpha s_{\bm r}^\gamma)\right\rangle_0=C_{\beta_Q}(\bm r)^2,
\label{eq:s_P_Q_exact_factorization}
\end{equation}
where $C_{\beta'}(\bm r)=\langle s_{\bm0}s_{\bm r}\rangle_{\beta'}$ is the spin correlation function of the clean Ising model at inverse temperature $\beta'$. 
Here $\langle\cdots\rangle_{\beta'}$ denotes the corresponding thermal expectation value.
The corresponding overlap susceptibilities are
\begin{equation}
    \chi_P^{(0)}(\bm k)=\sum_{\bm r}e^{i\bm k\cdot\bm r}C_\beta(\bm r)C_{\beta_Q}(\bm r),\qquad
    \chi_Q^{(0)}(\bm k)=\sum_{\bm r}e^{i\bm k\cdot\bm r}C_{\beta_Q}(\bm r)^2.
\label{eq:s_exact_overlap_chi}
\end{equation}
Expanding Eq.~\eqref{eq:s_lattice_effective_action} to quadratic order therefore gives
\begin{equation}
    S_{\rm eff}^{(2)}=\frac12\sum_\alpha\int_{\bm k}
    \left[\Lambda_P(\bm k)^{-1}-\chi_P^{(0)}(\bm k)\right]|\Psi_P^\alpha(\bm k)|^2
    +\frac12\sum_{\alpha<\gamma}\int_{\bm k}\left[\Lambda_Q(\bm k)^{-1}-\chi_Q^{(0)}(\bm k)\right]|\Psi_Q^{\alpha\gamma}(\bm k)|^2.
\label{eq:s_exact_RPA_kernel}
\end{equation}
where $\int_{\bm k}$ denotes integration over the first Brillouin zone with measure $\dd^dk/(2\pi)^d$.
Thus the lattice-HS quadratic kernels are
\begin{equation}
    \Gamma_P^{(2)}(\bm k)=\Lambda_P(\bm k)^{-1}-\chi_P^{(0)}(\bm k),\qquad
    \Gamma_Q^{(2)}(\bm k)=\Lambda_Q(\bm k)^{-1}-\chi_Q^{(0)}(\bm k).
\label{eq:s_exact_RPA_gamma}
\end{equation}
The corresponding RPA susceptibility is $\chi_X^{\rm RPA}=\chi_X^{(0)}/(1-\Lambda_X\chi_X^{(0)})$ for $X=P,Q$. 
The HS propagator and the RPA susceptibility have the same pole at 
\begin{equation}
    1-\Lambda_X(\bm k)\chi_X^{(0)}(\bm k)=0.
\label{eq:s_lattice_RPA_threshold}
\end{equation}


This formulation gives an immediate interpretation of noise mismatch. 
When $\beta=\beta_Q$, the true and posterior clean Ising models are identical, so $\chi_P^{(0)}(\bm k)=\chi_Q^{(0)}(\bm k)$.  
The relative enhancement of these two channels is determined entirely by
\begin{equation}
    \frac{\Lambda_Q(\bm k)}{\Lambda_P(\bm k)}
    =\frac{\lambda_Q}{\lambda_P}
    =\left(\frac{\Delta}{\Delta_Q}\right)^2.
\label{eq:s_lambda_ratio_noise}
\end{equation}
Hence
\begin{equation}
    \Delta_Q<\Delta\quad\Longrightarrow\quad
    \Lambda_Q\chi_Q^{(0)}>\Lambda_P\chi_P^{(0)},
\label{eq:s_noise_Q_favored}
\end{equation}
so underestimating the measurement noise immediately favors the posterior--posterior channel.

For temperature mismatch with $\Delta_Q=\Delta$, one instead has $\lambda_P=\lambda_Q=\Delta^{-2}$, and therefore $\Lambda_P(\bm k)=\Lambda_Q(\bm k)$.
The distinction between the two channels comes entirely from their clean-Ising susceptibilities, $\chi^{(0)}_X$. 
For the ferromagnetic Ising model, spin correlations are nonnegative and increase monotonically with inverse temperature. 
Thus for $\beta_Q>\beta$,
\begin{equation}
    C_{\beta_Q}(\bm r)\ge C_\beta(\bm r)\ge0.
\label{eq:s_corr_monotonic}
\end{equation}
Equation~\eqref{eq:s_exact_overlap_chi} then gives
\begin{equation}
    \chi_Q^{(0)}(0)-\chi_P^{(0)}(0)
    =\sum_{\bm r}C_{\beta_Q}(\bm r)
    \left[C_{\beta_Q}(\bm r)-C_\beta(\bm r)\right]\ge0.
\label{eq:s_chi_inequality}
\end{equation}
Hence a lower assumed temperature, $\beta_Q>\beta$, favors the posterior--posterior channel.
This difference directly controls the splitting of the two transitions along the temperature-mismatch cut. 
On a hypercubic lattice in $d$ dimension, $\Lambda_P(0)=\Lambda_Q(0)=2d/\Delta^2$, so the two instability conditions give
\begin{equation}
    \Delta_{c,P}^2=2d\,\chi_P^{(0)}(0),\qquad
    \Delta_{c,Q}^2=2d\,\chi_Q^{(0)}(0).
\label{eq:s_temperature_critical_delta}
\end{equation}
Therefore
\begin{equation}
    \Delta_{c,Q}^2-\Delta_{c,P}^2
    =2d\left[\chi_Q^{(0)}(0)-\chi_P^{(0)}(0)\right]\ge0.
\label{eq:s_temperature_transition_splitting}
\end{equation}
Within RPA, the clean-Ising susceptibility difference therefore directly determines the separation between the posterior-overlap and reconstruction transitions.

To determine how this splitting turns on near matching, write $\delta\beta=\beta_Q-\beta$ and consider the high-temperature regime $\beta, \beta_Q < \beta_c$. 
On a hypercubic lattice, the leading off-site correlations are the nearest-neighbor ones, $C_\beta(\bm r_{\rm NN})=\beta+O(\beta^3)$.
Since there are $2d$ nearest neighbors, Eq.~\eqref{eq:s_chi_inequality} gives
\begin{equation}
    \chi_Q^{(0)}(0)-\chi_P^{(0)}(0)
    =2d\left[\beta\,\delta\beta+(\delta\beta)^2\right]+\cdots.
\label{eq:s_chi_small_beta_dbeta}
\end{equation}
Consequently, for finite but small $\beta$ the splitting contains a term linear in $\delta\beta$ with a coefficient proportional to $\beta$, while at $\beta=0$ this linear contribution vanishes and the leading splitting is quadratic in $\beta_Q$.

At exact matching, $\beta_Q=\beta$ and $\Delta_Q=\Delta$, both the bare susceptibilities and the overlap couplings coincide.
The quadratic kernels therefore satisfy $\Gamma_P^{(2)}(\bm k)=\Gamma_Q^{(2)}(\bm k)$, as required by the enlarged $S_{n+1}$ replica symmetry.
At the matched critical point the two quadratic modes become critical simultaneously.
A mismatch breaks this enlarged symmetry and splits the two channels. 
Therefore, if a stable $Q$-only phase exists on the overconfident side, its boundary is expected to emanate from the matched multicritical point rather than requiring a finite mismatch. 

The same expansion also determines the leading cubic interactions. 
Since the one-point overlap averages vanish in the paramagnetic reference theory, the cubic cumulants reduce to the three-point contractions shown in Fig.~\ref{fig:Feynman_diagram}(b). 
For slowly varying overlap fields, the zero-momentum limits therefore generate the local vertices
\begin{equation}
    S_{\rm eff}^{(3)}=-w_1\int_{\bm r}\sum_{\alpha<\gamma}
    \Psi_P^\alpha\Psi_Q^{\alpha\gamma}\Psi_P^\gamma
    -w_2\int_{\bm r}\sum_{\alpha<\gamma<\delta}
    \Psi_Q^{\alpha\gamma}\Psi_Q^{\gamma\delta}\Psi_Q^{\alpha\delta},
\label{eq:s_lattice_cubic_action}
\end{equation}
with
\begin{equation}
    w_1=\int_{\bm k} C_\beta(\bm k)C_{\beta_Q}(\bm k)^2,\qquad
    w_2=\int_{\bm k} C_{\beta_Q}(\bm k)^3,
\label{eq:s_lattice_cubic}
\end{equation}
where $C_{\beta'}(\bm k)$ is the Fourier transform of the clean-Ising correlation function.
Thus the lattice formulation expresses both the quadratic kernels and the leading nonlinear vertices directly in terms of clean-Ising correlations. 
At matching, $\beta_Q=\beta$, one has $w_1=w_2$, as required by the enlarged $S_{n+1}$ symmetry.

\subsection{Universality classes of the phase transitions}
\label{sec:s_transition_rg}

The continuum replica field theory is 
\begin{align}
    S_{\rm eff}={}&\frac12\int_{\bm r}\left[\sum_\alpha [(\nabla\Psi_P^\alpha)^2 + r_P (\Psi_P^\alpha)^2] + \sum_{\alpha<\gamma}[(\nabla\Psi_Q^{\alpha\gamma})^2 + r_Q (\Psi_Q^{\alpha\gamma})^2]\right]\label{seq:exact_RPA_kernel}\\
    &-\int_{\bm r}\left[w_1\sum_{\alpha<\gamma}\Psi_P^\alpha\Psi_Q^{\alpha\gamma}\Psi_P^\gamma+w_2\sum_{\alpha<\gamma<\delta}\Psi_Q^{\alpha\gamma}\Psi_Q^{\gamma\delta}\Psi_Q^{\alpha\delta}\right].\nonumber
\end{align}
The mass terms, $r_P$ and $r_Q$, are independent. 
As discussed above, the uninformative phase is realized at $r_P,r_Q>0$, while $r_P < 0$ realizes the informative phase, and $r_Q <  0 < r_P$ realizes the overconfident phase.
We analyze the universality of the transitions between these phases. 
The symmetry of the replicated theory is
\begin{equation}
    G=Z_2^{(0)}\times\left[(Z_2)^n\rtimes S_n\right],
\label{eq:s_full_replica_symmetry}
\end{equation}
where $Z_2^{(0)}$ acts on the true configuration and $(Z_2)^n$ denotes the independent Ising transformations of the posterior replicas. 
The semidirect product indicates that $S_n$ permutes these $n$ posterior $Z_2$ factors.

In the uninformative phase, neither overlap field condenses,
\begin{equation}
    \langle\Psi_P^\alpha\rangle=0,\qquad \langle\Psi_Q^{\alpha\gamma}\rangle=0,
\label{eq:s_uninformative_order}
\end{equation}
so the full symmetry in Eq.~\eqref{eq:s_full_replica_symmetry} remains unbroken.
In the overconfident phase, posterior replicas develop mutual long-range overlap while reconstruction remains absent.
Within a replica-symmetric description,
\begin{equation}
    \langle\Psi_P^\alpha\rangle=0,\qquad
    \langle\Psi_Q^{\alpha\gamma}\rangle=q\neq0,\qquad \alpha\neq\gamma.
\label{eq:s_overconfident_order}
\end{equation}
Under independent global $ Z_2$ flips of the posterior replicas, $\Psi_Q^{\alpha\gamma}\to \epsilon_\alpha\epsilon_\gamma\Psi_Q^{\alpha\gamma}$, with $\epsilon_\alpha=\pm1$. 
A replica-symmetric condensate $\langle\Psi_Q^{\alpha\gamma}\rangle\neq0$ is therefore invariant only if $\epsilon_1=\cdots=\epsilon_n$. 
Thus the independent replica symmetry $(\mathbb Z_2)^n$ is broken down to the diagonal $\mathbb Z_2$.
The true replica remains independently flippable. 
The unbroken symmetry is therefore
\begin{equation}
    G_{\rm over}=Z_2^{(0)}\times Z_2^{(Q)}\times S_n,
\label{eq:s_overconfident_symmetry}
\end{equation}
where $Z_2^{(Q)}$ denotes the simultaneous global flip of all posterior replicas.

In the informative phase, the truth--posterior overlap also becomes nonzero,
\begin{equation}
    \langle\Psi_P^\alpha\rangle=p\neq0.
\label{eq:s_informative_order}
\end{equation}
Because $\Psi_P^\alpha$ transforms as $\epsilon_0\epsilon_\alpha\Psi_P^\alpha$, invariance requires the true and posterior replicas to transform together, $\epsilon_\alpha=\epsilon_0$ for all $\alpha$. 
The independent true and posterior $Z_2$ transformations are therefore locked into a single diagonal Ising symmetry,
\begin{equation}
    G_{\rm info}=Z_2^{\rm diag}\times S_n.
\label{eq:s_informative_symmetry}
\end{equation}
The three learning phases can thus be organized by the symmetry-breaking sequence
\begin{equation}
    Z_2^{(0)}\times\left[(Z_2)^n\rtimes S_n\right]
    \longrightarrow
    Z_2^{(0)}\times Z_2^{(Q)}\times S_n
    \longrightarrow
    Z_2^{\rm diag}\times S_n.
\label{eq:s_learning_symmetry_hierarchy}
\end{equation}
The first breaking corresponds to the onset of posterior--posterior agreement without reconstruction, while the second locks the true configuration to the posterior replicas and produces an informative phase. 
This symmetry hierarchy is intrinsic to the replicated inference problem. 
Under the RBIM mapping discussed in the next section, these three phases correspond respectively to the paramagnetic (PM), spin-glass (SG), and ferromagnetic (FM) regimes of the associated random-bond Ising model, which allows their critical behavior to be connected to the corresponding known universality classes.

\subsubsection{Nishimori multicritical point}

At Bayes matching, the true configuration becomes statistically equivalent to the posterior replicas, and the permutation symmetry is enlarged from $S_n$ to $S_{n+1}$.
The two overlap fields can be combined into a single symmetric matrix,
\begin{equation}
    \Psi^{0\alpha}=\Psi_P^\alpha,\qquad \Psi^{\alpha\gamma}=\Psi_Q^{\alpha\gamma},\qquad \Psi^{aa}=0,
\label{eq:s_transition_full_overlap}
\end{equation}
with replica indices $a,b=0,\ldots,n$. The enlarged symmetry is
\begin{equation}
    G_{\rm N}=(Z_2)^{n+1}\rtimes S_{n+1}.
\label{eq:s_nishimori_symmetry}
\end{equation}

At the multicritical point, the reconstruction and posterior-overlap sectors become critical simultaneously.
The leading nonlinear invariant allowed by the enlarged symmetry is cubic,
\begin{equation}
    \mathcal L_{\rm N}
    =\frac12\sum_{a< b}\left[(\nabla\Psi^{ab})^2+r(\Psi^{ab})^2\right]
    -w\sum_{a<b<c}\Psi^{ab}\Psi^{bc}\Psi^{ac}+\cdots .
\label{eq:s_nishimori_field_theory}
\end{equation}
The upper critical dimension is $d_c=6$. 
The corresponding $6-\epsilon$ RG was developed in Refs.~\cite{Chen1977Mean,Doussal1989epsilon}.
In the replica limit, the one-loop multicritical fixed point lies on the Nishimori manifold and gives
\begin{equation}
    \nu=\frac12+\frac{5\epsilon}{12}+O(\epsilon^2), \qquad \eta=-\frac{\epsilon}{3}+O(\epsilon^2).
\label{eq:s_nishimori_exponents}
\end{equation}
More recently, Ref.~\cite{wiese2026bayesian} extended the RG analysis to two-loop order for general replica number $n_{\rm tot}$.
Here, $n_{\rm tot}$ refers to the dimension of the matrix field $\Psi$ in the Lagrangian Eq.~\eqref{eq:s_nishimori_field_theory}. 
For the symmetry-preserving scaling direction, they obtain
\begin{equation}
    \frac{1}{\nu}=2-\frac{5}{3}\epsilon+\frac{973-686(n_{\rm tot}-1)+73(n_{\rm tot}-1)^2}{54(n_{\rm tot}-2)^2}\epsilon^2+O(\epsilon^3),\qquad
    \eta=-\frac{\epsilon}{3}+\frac{88-2(n_{\rm tot}-1)(n_{\rm tot}+24)}{27(n_{\rm tot}-2)^2}\epsilon^2+O(\epsilon^3).
\end{equation}
In their convention, the Nishimori multicritical point corresponds to $n_{\rm tot}\to1$, giving
\begin{equation}
    \frac{1}{\nu}=2-\frac{5}{3}\epsilon+\frac{973}{54}\epsilon^2+O(\epsilon^3),
    \qquad
    \eta=-\frac{\epsilon}{3}+\frac{88}{27}\epsilon^2+O(\epsilon^3).
\label{eq:s_nishimori_exponents_two_loop}
\end{equation}
The Nishimori multicritical behavior and its two relevant scaling directions have also been studied numerically in two and three dimensions~\cite{Honecker2001Universality,Hasenbusch2008Nishimori2D,deQueiroz2009Location,Hasenbusch2007Magnetic}.

\subsubsection{Uninformative--informative transition}

We next consider the direct transition between the uninformative and informative phases. 
Along this transition, the reconstruction field $\Psi_P^\alpha$ becomes critical while the posterior-overlap field $\Psi_Q^{\alpha\gamma}$ remains noncritical.
The associated symmetry breaking is
\begin{equation}
    Z_2^{(0)}\times\left[(Z_2)^n\rtimes S_n\right]
    \longrightarrow
    Z_2^{\rm diag}\times S_n.
\label{eq:s_uninfo_info_symmetry}
\end{equation}

Because the $\Psi_Q$ sector remains massive, it can be integrated out at long distances. 
The cubic coupling $\Psi_P^\alpha\Psi_Q^{\alpha\gamma}\Psi_P^\gamma$ then generates an inter-replica quartic interaction for the critical reconstruction field.
Together with the local quartic interaction, this yields the standard replicated random-bond Ising field theory,
\begin{equation}
    \mathcal L_{\rm uninfo-info}
    =\frac12\sum_\alpha\left[(\nabla \Psi_P^\alpha)^2+r_P (\Psi_P^\alpha)^2\right]
    +\frac{u}{4!}\sum_\alpha (\Psi_P^\alpha)^4
    -\frac{v}{4!}\sum_{\alpha,\gamma} (\Psi_P^\alpha)^2 (\Psi_P^\gamma)^2.
\label{eq:s_uninfo_info_random_mass}
\end{equation}
The quartic interactions are marginal at $d_c=4$.

The RG of the replicated random-bond Ising theory has been studied before~\cite{Grinstein1976Application}. 
In the replica limit $n\rightarrow0$, the leading solution of the two beta functions becomes degenerate, so the random fixed point is organized in powers of $\sqrt{\epsilon}$ rather than $\epsilon$, with $\epsilon=4-d$.
The leading critical exponents are
\begin{equation}
    \nu=\frac12+\frac14\sqrt{\frac{6\epsilon}{53}}+O(\epsilon),\qquad
    \eta=-\frac{\epsilon}{106}+O(\epsilon^{3/2}).
\label{eq:s_uninfo_info_epsilon}
\end{equation}
The $\sqrt{\epsilon}$ expansion has been extended to six-loop order in Ref.~\cite{Kompaniets2021Critical}.
Thus the direct uninformative--informative transition is controlled by the random-bond Ising universality class rather than by the cubic $6-\epsilon$ theory governing the Nishimori multicritical point.
Numerical studies of the three-dimensional RBIM confirm this identification with the randomly dilute Ising universality class~\cite{Hasenbusch2007Critical,Hasenbusch2007universality}.

\subsubsection{Uninformative--overconfident transition}

We next consider the transition from the uninformative phase to the overconfident phase, which corresponds to the PM--SG transition in the statistical model mapping. 
Along this transition, the reconstruction field $\Psi_P^\alpha$ remains massive while the posterior-overlap field $\Psi_Q^{\alpha\gamma}$ becomes critical.
The associated symmetry breaking is
\begin{equation}
    Z_2^{(0)}\times\left[(Z_2)^n\rtimes S_n\right]
    \longrightarrow
    Z_2^{(0)}\times Z_2^{(Q)}\times S_n.
\label{eq:s_uninfo_over_symmetry}
\end{equation}

Since the $P$ sector remains noncritical, it can be integrated out at long distances. 
The critical theory is therefore expressed solely in terms of the posterior-overlap field,
\begin{equation}
    \mathcal L_{\rm uninfo-over}
    =\frac12\sum_{\alpha<\gamma}
    \left[(\nabla \Psi_Q^{\alpha\gamma})^2+r_Q (\Psi_Q^{\alpha\gamma})^2\right]
    -w\sum_{\alpha<\gamma<\delta}
    \Psi_Q^{\alpha\gamma} \Psi_Q^{\gamma\delta} \Psi_Q^{\alpha\delta}.
\label{eq:s_uninfo_over_action}
\end{equation}
The replica limit is $n\rightarrow0$. 
The upper critical dimension is $d_c=6$ for the cubic coupling.

The RG of this cubic replica theory is the conventional field theory of the zero-field Ising spin-glass transition~\cite{Chen1977Mean,Doussal1989epsilon}. 
In $d=6-\epsilon$, the leading critical exponents in the Ising case are
\begin{equation}
    \nu=\frac12+\frac{5\epsilon}{12}+O(\epsilon^2), \qquad \eta=-\frac{\epsilon}{3}+O(\epsilon^2).
\label{eq:s_uninfo_over_epsilon}
\end{equation}

The higher-order expansion was obtained for the $m$-component vector spin glass in Ref.~\cite{Green1985expansion}.
Keeping terms through $O(\epsilon^2)$, the result is
\begin{align}
    \eta&=-\frac{1}{3}\frac{m\epsilon}{2m-1}
    +\frac{\epsilon^2}{4^3(2m-1)^3}
    \left(-\frac{104}{9}m+\frac{2752}{27}m^2-\frac{88}{9}m^3\right)
    +O(\epsilon^3),
    \\
    \nu^{-1}-2+\eta&=-\frac{2m}{2m-1}\epsilon
    +\frac{\epsilon^2}{4^3(2m-1)^3}
    \left(-48m+\frac{1312}{3}m^2+\frac{1840}{9}m^3\right)
    +O(\epsilon^3).
\end{align}
The explicit $O(\epsilon^3)$ contributions were also obtained in Ref.~\cite{Green1985expansion}, but are not displayed here.
In their convention, the uninformative--overconfident transition corresponds to the replica limit $n\to0$, with the same form of the action as in Eq.~\eqref{eq:s_uninfo_over_action}.
In particular, three-dimensional numerical studies find substantially different critical behavior at the ordinary spin-glass transition and at the Nishimori multicritical point~\cite{Hasenbusch2007Magnetic,Hasenbusch2007Critical,Baity2013Critical}.

It is useful to compare this theory with the Nishimori multicritical theory discussed above. 
At the Nishimori point, the true replica joins the posterior replicas, and the fields $\Psi_P^\alpha=\Psi^{0\alpha}$ and $\Psi_Q^{\alpha\gamma}=\Psi^{\alpha\gamma}$ form a single overlap matrix with $n+1$ replica indices.
Since the posterior replica limit remains $n\rightarrow0$, the fully symmetric Nishimori theory can equivalently be viewed as the same cubic overlap theory analytically continued to a total replica number $n_{\rm tot}\rightarrow1$.
By contrast, the uninformative--overconfident transition is the conventional spin-glass theory with $n_{\rm tot}\rightarrow0$.

A closely related replica theory arises for Bayesian inference of the critical Ising model~\cite{wiese2026bayesian}.
There, the single-replica Ising field and the overlap field combine into an enlarged symmetric overlap matrix with $n_{\rm tot}$ indices. 
In terms of the total number of overlap indices, the three problems therefore correspond to the following replica limits, 
\begin{equation}
    \begin{split}
        \text{uninformative--overconfident}:&\qquad n_{\rm tot}\rightarrow0,\\
        \text{Nishimori}:&\qquad n_{\rm tot}\rightarrow1,\\
        \text{critical-Ising optimal inference}:&\qquad n_{\rm tot}\rightarrow2.
    \end{split}
\label{eq:s_replica_limits_sg_nishimori}
\end{equation}

This relation also clarifies the perturbative results quoted above.
For $n_{\rm tot}<2$, Ref.~\cite{wiese2026bayesian} gives the two-loop exponents for general replica number.
At $n_{\rm tot}=0$, corresponding to the ordinary PM--SG transition, their result reduces to
\begin{equation}
    \frac{1}{\nu}=2-\frac{5}{3}\epsilon+\frac{433}{54}\epsilon^2+O(\epsilon^3),\qquad
    \eta=-\frac{\epsilon}{3}+\frac{34}{27}\epsilon^2+O(\epsilon^3),
\end{equation}
in exact agreement through $O(\epsilon^2)$ with the $m=1$ specialization of Ref.~\cite{Green1985expansion}, which further gives the explicit $O(\epsilon^3)$ corrections for this conventional spin-glass replica limit.
For the Nishimori problem, $n_{\rm tot}=1$, the result instead gives Eq.~\eqref{eq:s_nishimori_exponents_two_loop}. 
The $O(\epsilon)$ terms are independent of $n_{\rm tot}$, explaining why the Nishimori and PM--SG transitions have the same leading expressions, while their replica-number dependence first appears at $O(\epsilon^2)$.

The critical-Ising case, $n_{\rm tot}\rightarrow2$, is qualitatively different.
The general $n_{\rm tot}<2$ fixed point and critical exponents are singular in this limit, and therefore the critical-Ising exponents cannot be obtained by simply setting $n_{\rm tot}=2$ in the general-replica formulas.
Indeed, the one-loop contribution that stabilizes the cubic fixed point for $n{\rm tot}<2$ vanishes at $n_{\rm tot}=2$, so that the two-loop RG terms become essential already at leading order.
This singular replica limit was treated separately in Ref.~\cite{wiese2026bayesian}, leading to a distinct perturbative expansion for the optimal inference transition of the critical Ising model.

\subsubsection{Overconfident--informative transition}

We finally consider the transition from the overconfident phase to the informative phase, corresponding to the SG--FM transition in the statistical model.
The posterior-overlap field is already ordered on both sides of the transition, while the reconstruction field changes from $\langle\Psi_P^\alpha\rangle=0$ to $\langle\Psi_P^\alpha\rangle\neq0$.
The associated symmetry breaking is
\begin{equation}
    Z_2^{(0)}\times Z_2^{(Q)}\times S_n
    \longrightarrow
    Z_2^{\rm diag}\times S_n.
\end{equation}
Thus only the relative $Z_2$ symmetry between the true replica and the posterior replicas is broken.

For simplicity, we take the ordered posterior-overlap field to be replica symmetric and neglect its fluctuations, $Q_{\alpha\gamma}=q$ for $\alpha\neq\gamma$.
The cubic coupling between the reconstruction and overlap sectors then gives
\begin{equation}
    -w_1q\sum_{\alpha<\gamma}\Psi_P^\alpha\Psi_P^\gamma
    =-\frac{w_1q}{2}\left[\left(\sum_\alpha\Psi_P^\alpha\right)^2-\sum_\alpha(\Psi_P^\alpha)^2\right].
\end{equation}
The second term can be absorbed into the diagonal mass, denoted by $t$. 
Defining $\mathcal D=w_1q$, the quadratic theory therefore takes the same form as the replicated random-field Ising model.

Once the overlap background is fixed, the remaining symmetry is only $Z_2^{(0)}\times Z_2^{(Q)}\times S_n$. 
The most general quartic potential invariant under this symmetry contains five independent $S_n$ singlets~\cite{Kaviraj2021Random},
\begin{equation}
    I_1=\sum_\alpha (\Psi_P^\alpha)^4,\ \
    I_2=\left(\sum_\alpha \Psi_P^\alpha\right)\left(\sum_\alpha (\Psi_P^\alpha)^3\right),\ \
    I_3=\left(\sum_\alpha (\Psi_P^\alpha)^2\right)^2,\ \
    I_4=\left(\sum_\alpha \Psi_P^\alpha\right)^2\left(\sum_\alpha (\Psi_P^\alpha)^2\right),\ \
    I_5=\left(\sum_\alpha \Psi_P^\alpha\right)^4.
\end{equation}
The corresponding effective theory can therefore be written as
\begin{equation}
    \mathcal L_{\rm over-info}^{\rm RS}
    =\sum_{\alpha=1}^n\left[\frac12(\nabla \Psi_P^\alpha)^2+\frac t2 (\Psi_P^\alpha)^2\right]
    -\frac{\mathcal D}{2}\left(\sum_\alpha \Psi_P^\alpha\right)^2+\frac{u}{4!}I_1+\frac{v_1}{4!}I_2+\frac{v_2}{4!}I_3+\frac{v_3}{4!}I_4+\frac{v_4}{4!}I_5+\cdots .
\end{equation}
The $I_1$ and $I_3$ interactions are already allowed by the larger $(Z_2)^n\rtimes S_n$ symmetry before the overlap field orders. 
The remaining terms become symmetry allowed after choosing an ordered spin-glass background and can in general be generated in the effective theory after the overlap sector is integrated out.

The power counting is most transparent in the Cardy basis~\cite{Kaviraj2021Random},
\begin{equation}
    \Psi_P^1=\varphi+\frac{\omega}{2},\qquad
    \Psi_P^\alpha=\varphi-\frac{\omega}{2}+\chi_\alpha,\qquad
    \sum_{\alpha=2}^n\chi_\alpha=0,
\end{equation}
for which the Gaussian scaling dimensions are
\begin{equation}
    [\varphi]=\frac d2-2,\qquad
    [\omega]=\frac d2,\qquad
    [\chi]=\frac d2-1.
\label{eq:cardy_dimension}
\end{equation}
In the replica limit,
\begin{equation}
    \sum_\alpha \Psi_P^\alpha\xrightarrow{n\to0}\omega,\qquad
    \sum_\alpha (\Psi_P^\alpha)^2\xrightarrow{n\to0}2\varphi\omega+\sum_{\alpha=2}^n\chi_\alpha^2,\qquad
    \sum_\alpha (\Psi_P^\alpha)^3\xrightarrow{n\to0}
    3\varphi^2\omega+3\varphi\sum_{\alpha=2}^n\chi_\alpha^2+\cdots .
\label{eq:expansion}
\end{equation}
The usual single-replica quartic interaction contains
\begin{equation}
    I_1=\sum_{\alpha=1}^n (\Psi_P^\alpha)^4
    \xrightarrow{n\to0}
    4\omega\varphi^3+6\varphi^2\sum_{\alpha=2}^n\chi_\alpha^2+\cdots ,
\end{equation}
whose leading operators have scaling dimension $2d-6$.
The corresponding coupling therefore has Gaussian RG eigenvalue $y_u=6-d$ and becomes marginal at $d_c=6$, reproducing the upper critical dimension of the random field Ising model (RFIM).

The remaining quartic singlets have higher Gaussian scaling dimensions. 
Using Eqs.~\eqref{eq:cardy_dimension} and~\eqref{eq:expansion}, their leading dimensions are
\begin{equation}
    \Delta_{I_2}=2d-4,\qquad
    \Delta_{I_3}=2d-4,\qquad
    \Delta_{I_4}=2d-2,\qquad
    \Delta_{I_5}=2d.
\end{equation}
Consequently, near $d=6-\epsilon$, only $I_1$ is perturbatively relevant close to six dimensions, while all additional quartic interactions are irrelevant.
The leading critical theory in the controlled $6-\epsilon$ expansion is therefore the RFIM.

The perturbative RFIM fixed point obeys dimensional reduction.
Its critical exponents in $d$ dimensions coincide order by order in perturbation theory with those of the pure Ising model in $d-2$ dimensions~\cite{Aharony1976Lowering,Parisi1979Random,Kaviraj2021Random}. 
Hence, for $d=6-\epsilon$, $\nu_{\rm RFIM}\big|_{d=6-\epsilon}=\nu_{\rm Ising}\big|_{d=4-\epsilon}$, and the Wilson--Fisher result gives
\begin{equation}
    \frac{1}{\nu}=2-\frac{\epsilon}{3}+O(\epsilon^2).
\end{equation}

The situation can change as the spatial dimension is lowered.
In particular, $I_2$ and $I_3$ become marginal by Gaussian power counting at $d=4$ and are formally relevant below four dimensions, whereas $I_4$ and $I_5$ remain irrelevant in $d=3$.
Therefore the truncation to the RFIM is controlled near $d=6$, but is no longer protected by simple power counting upon continuation to three dimensions.
More generally, additional $S_n$-invariant perturbations of the replicated RFIM are known to become important as the dimension is lowered~\cite{Kaviraj2021Random}. 
The overconfident--informative transition may consequently flow away from the perturbative RFIM fixed point and approach a different strong-coupling universality class in three dimensions.
Determining the resulting fixed point requires an RG treatment beyond the leading $6-\epsilon$ expansion.


\subsubsection{Summary}

Table~\ref{tab:s_rbim_exponents} summarizes the critical exponents obtained numerically in three dimensions together with the leading perturbative RG predictions discussed above.
The numerical estimates refer to the corresponding three-dimensional RBIM universality classes, while the analytical results are expansions about their respective upper critical dimensions and should not in general be quantitatively extrapolated to $d=3$.

\begin{table}[t]
    \centering
    \caption{Critical exponents for the various inference transitions, including numerical estimates in $d=3$ and the leading perturbative RG predictions. 
    For the Nishimori and PM--SG transitions, $\epsilon=6-d$, while for the PM--FM transition $\epsilon=4-d$.
    The FM--SG analytical entry denotes the RFIM prediction near $d=6$ rather than an established RG result for the three-dimensional ferromagnetic--glassy fixed point.}
\label{tab:s_rbim_exponents}
\renewcommand{\arraystretch}{2.2}
\begin{tabular}{c|c|c}
\hline\hline
Transition & Numerical, $d=3$ & Perturbative RG \\
\hline
Optimal (Nishimori)
&
$\nu=0.98(5),\ \eta=-0.114(3)$~\cite{Hasenbusch2007Magnetic}
&
$\nu=\dfrac12+\dfrac{5}{12}\epsilon+O(\epsilon^2),\ \eta=-\dfrac{\epsilon}{3}+O(\epsilon^2)$~\cite{Chen1977Mean,Doussal1989epsilon,wiese2026bayesian,Green1985expansion}
\\
\hline
Uninformative--Informative (PM--FM)
&
$\nu=0.682(3),\ \eta=0.036(2)$~\cite{Hasenbusch2007Critical}
&
$\nu=\dfrac12+\dfrac14\sqrt{\dfrac{6}{53}}\,\epsilon^{1/2}+O(\epsilon),\ \eta=-\dfrac{\epsilon}{106}+O(\epsilon^{3/2})$~\cite{Grinstein1976Application,Kompaniets2021Critical}
\\
\hline
Uninformative--Overconfident (PM--SG)
&
$\nu=2.562(42),\ \eta=-0.3900(36)$~\cite{Baity2013Critical}
&
$\nu=\dfrac12+\dfrac{5}{12}\epsilon+O(\epsilon^2),\ \eta=-\dfrac{\epsilon}{3}+O(\epsilon^2)$~\cite{Harris1976Critical,Chen1977Mean,Doussal1989epsilon,wiese2026bayesian,Green1985expansion}
\\
\hline
Informative--Overconfident (FM--SG)
&
$\nu=0.96(2),\ \eta=-0.39(2)$~\cite{Ceccarelli2011Ferromagnetic}
&
$\dfrac{1}{\nu}=2-\dfrac{\epsilon}{3}+O(\epsilon^2),\ \eta=O(\epsilon^2)$~\cite{Kaviraj2021Random}
\\
\hline\hline
\end{tabular}
\end{table}


\subsection{Replica field theory for mismatched Bayesian inference for the critical Ising model}

The discussion above focused on the regime $\beta,\beta_Q<\beta_c$, where the individual true and posterior Ising models are in their paramagnetic phases.
Their single-replica magnetic fluctuations are then massive, so the long-distance theory can be formulated entirely in terms of the overlap fields.
In the Bayesian inference for the critical Ising model with $\beta = \beta_Q = \beta_c$, we need to retain the coarse-grained spin fields of the individual replicas together with the replica-overlap fields. 
At criticality, the Ising model is described by a critical scalar field, denoted as $\phi_a$, for $a=0,...,n$. 
Including the critical scalar field for each replica, the coupled continuum theory takes the form
\begin{eqnarray}
    \mathcal L &=& \frac12 \sum_{a}(\nabla \phi^a)^2 + \frac12 \sum_{\alpha}[(\nabla \Psi_P^{\alpha})^2 + r_P (\Psi_P^{\alpha})^2] +  \frac12 \sum_{\alpha<\beta} [(\nabla \Psi_Q^{\alpha\beta})^2 + r_Q (\Psi_Q^{\alpha\beta})^2 ] \nonumber \\
    && + g_1 \sum_{\alpha} \Psi_P^{\alpha} \phi^0 \phi^\alpha + g_2 \sum_{\alpha<\beta} \Psi_Q^{\alpha\beta} \phi^\alpha \phi^\beta + w_1 \sum_{\alpha<\beta} \Psi_P^{\alpha} \Psi_Q^{\alpha\beta} \Psi_P^{\beta} + w_2 \sum_{\alpha<\beta<\gamma}\Psi_Q^{\alpha\beta} \Psi_Q^{\beta\gamma} \Psi_Q^{\gamma\alpha},
\end{eqnarray}
where $g_1$ and $g_2$ denote the new coupling to the fluctuation from the critical Ising model, which is described by a Gaussian model at $d>4$. 
We expect the symmetry and the phase diagram remain the same.
However, the coupling to the critical Ising model will modify the universality class.
The matched Bayesian inference for the critical Ising model was studied in~\cite{patil2026higher,wiese2026bayesian}

Here we consider the simplest mismatched extension, with $\beta=\beta_Q=\beta_c$ and $\Delta\neq\Delta_Q$.
The measurement mismatch changes the microscopic overlap interactions, with $\lambda_P=\Delta_Q^{-2},\lambda_Q=\Delta^2/{\Delta_Q^4}$.
After the Hubbard--Stratonovich transformation, these couplings enter inversely in the quadratic kernels of $\Psi_P$ and $\Psi_Q$.
Since $\Lambda_X(0)=2d\lambda_X$, the bare mass splitting is, up to a common positive normalization,
\begin{equation}
    r_Q-r_P
    \propto
    \frac{1}{\lambda_Q}-\frac{1}{\lambda_P}
    =
    \Delta_Q^2
    \left(
        \frac{\Delta_Q^2}{\Delta^2}-1
    \right).
\end{equation}
Therefore $\Delta_Q<\Delta$ makes the posterior--posterior sector softer than the truth--posterior sector, $r_Q<r_P$, providing the same tendency toward a separated overconfident regime as in the noncritical case.
The phase hierarchy also remains unchanged, since the cubic coupling $w_1\Psi_P^\alpha\Psi_Q^{\alpha\beta}\Psi_P^\beta$ implies that condensation of $\Psi_P$ induces $\Psi_Q$.
Thus, we expect that the coupling to the critical Ising fields modifies the universality classes of these transitions, but not the mismatch-induced splitting or this ordering structure.

\section{Numerical methods and three-dimensional results}
\label{sec:s_numerics}

In the RBIM representation, the reconstruction and posterior-agreement channels become ferromagnetic (FM) and spin-glass (SG) observables. 
We first define the finite-size observables and Monte Carlo procedure, and present the additional crossing data underlying the phase diagram in the main text.

\subsection{Finite-size observables}
\label{sec:s_3d_observables}

We consider periodic simple-cubic lattices with $N_s=L^3$ sites. 
For each quenched bond realization, two conditionally independent relative-spin replicas $\sigma^1$ and $\sigma^2$ are simulated with the same couplings $J_{ij}$. 
We define
\begin{equation}
    p_i=\sigma_i,\qquad q_i=\sigma_i^1\sigma_i^2.
\label{eq:s_numerical_fields}
\end{equation}
In the original learning variables, these correspond to the truth--posterior overlap $p_i=s_i^0s_i^1$ and the posterior--posterior overlap $q_i=s_i^1s_i^2$.
Their structure factors are
\begin{equation}
    \chi_P(\bm k)=\frac1{N_s}\E\left\langle\left|\sum_jp_je^{i\bm k\cdot\bm r_j}\right|^2\right\rangle,\qquad
    \chi_Q(\bm k)=\frac1{N_s}\E\left\langle\left|\sum_jq_je^{i\bm k\cdot\bm r_j}\right|^2\right\rangle.
\label{eq:s_structure_factors}
\end{equation}
The outer average is over the induced quenched disorder, equivalently over independent realizations of the true configuration and Gaussian measurement noise.
The correlation lengths are
\begin{equation}
    \xi_X=\frac1{2\sin(\pi/L)}\left[\frac{\chi_X(\bm0)}{\chi_X(\bm k_{\min})}-1\right]^{1/2}, \qquad X=P,Q.
\label{eq:s_xi}
\end{equation}
We average $\chi_X(\bm k_{\min})$ over symmetry-related momenta with one component $2\pi/L$ and all other components zero. 
The dimensionless ratios $\xi_P/L$ and $\xi_Q/L$ provide the primary finite-size diagnostics. 
Crossings of $\xi_P/L$ locate the FM transition, while crossings of $\xi_Q/L$ locate the SG transition.

\subsection{Monte Carlo updates and replica exchange}
\label{sec:s_mc_updates}

For a fixed bond realization, the relative spins are updated using local heat-bath dynamics. 
The local field acting on $\sigma_i$ is $h_i=\sum_{j\in\partial i}J_{ij}\sigma_j$.
The conditional probability for $\sigma_i=+1$ is
\begin{equation}
    P(\sigma_i=+1|\{\sigma_{j\ne i}\})=\frac1{1+e^{-2h_i}}.
\label{eq:s_heatbath}
\end{equation}
Even and odd sublattices are updated alternately, and two statistically independent relative-spin replicas are evolved for each quenched realization in order to measure the SG observables.

To improve equilibration in the glassy regime, we supplement the local updates with replica exchange between neighboring points of the scanned parameter ladder~\cite{hukushima1996exchange}. 
Let $\sigma_d$ and $\sigma_{d'}$ denote configurations at neighboring ladder points $d$ and $d'$, with corresponding couplings $J_{ij}^{(d)}$ and $J_{ij}^{(d')}$.
Exchanging the two configurations is accepted with probability $P_{\rm acc}=\min\{1,e^{\log R}\}$, where
\begin{equation}
    \log R=\sum_{\langle ij\rangle}\left(J_{ij}^{(d)}-J_{ij}^{(d')}\right)
    \left(\sigma_{i,d'}\sigma_{j,d'}-\sigma_{i,d}\sigma_{j,d}\right).
\label{eq:s_exchange_ratio}
\end{equation}
Exchange attempts are alternated between even and odd neighboring pairs of ladder points after successive Monte Carlo sweeps.
The total thermalization sweeps for all parameter scans and system sizes are summarized in Table~\ref{tab:thermalization_sweeps}.

\begin{table}[t]
    \centering
    \caption{Thermalization sweeps for the three-dimensional Monte Carlo simulations.
    Each entry gives the total number of equilibration sweeps at the final accepted thermalization check.}
    \begin{ruledtabular}
    \begin{tabular}{lrrrr}
    Parameter scan & $L=4$ & $L=6$ & $L=8$ & $L=10$ \\
    \colrule
    {$\beta=\beta_Q=0$, $\Delta=2$, varying $\Delta_Q$}
    & 6000 & 10000 & 24000 & 104000\\
    $\beta=\beta_Q=0$, $\Delta/\Delta_Q=0.25$, varying $\Delta_Q$
    & 6000 & 10000 & 20000 & 40000 \\
    $\beta=\beta_Q=0$, $\Delta/\Delta_Q=0.5$, varying $\Delta_Q$
    & 6000 & 18000 & 20000 & 40000 \\
    $\beta=\beta_Q=0$, $\Delta/\Delta_Q=0.7$, varying $\Delta_Q$
    & 6000 & 18000 & 20000 & 40000 \\
    $\beta=\beta_Q=0$, $\Delta/\Delta_Q=1.0$, varying $\Delta_Q$
    & 6000 & 10000 & 20000 & 48000 \\
    $\beta=\beta_Q=0$, $\Delta/\Delta_Q=1.3$, varying $\Delta_Q$
    & 8000 & 32000 & 112000 & 408000 \\
    $\beta=0.011$, $\beta_Q=0.002$, $\Delta=\Delta_Q$, varying $\Delta$
    & 6000 & 14000 & 28000 & 60000\\
    $\beta=0.011$, $\beta_Q=0.011$, $\Delta=\Delta_Q$, varying $\Delta$
    & 6000 & 14000 & 28000 & 78000\\
    $\beta=0.011$, $\beta_Q=0.14$, $\Delta=\Delta_Q$, varying $\Delta$
    & 6000 & 12000 & 32000 & 78000\\
    $\beta=0.011$, $\beta_Q=0.22$, $\Delta=\Delta_Q$, varying $\Delta$
    & 10000 & 18000 & 40000 & 108000
    \end{tabular}
    \end{ruledtabular}
    \label{tab:thermalization_sweeps}
\end{table}

\subsection{Additional three-dimensional finite-size data}
\label{sec:s_3d_phase_boundaries}

We present here the additional finite-size data underlying the noise-mismatch phase diagram in Fig.~\ref{fig:phase}(a) of the main text. 
All calculations use periodic simple-cubic lattices with $\beta=\beta_Q=0$.

Figure~\ref{fig:transitionD} shows scans in which $\Delta_Q$ is varied at fixed ratios $\Delta/\Delta_Q=1,0.7,0.5,0.25$. 
These crossings determine the corresponding FM--PM transition points. On the Nishimori line, $\Delta=\Delta_Q$, the ferromagnetic and spin-glass correlations are identical after disorder averaging, and the crossings of $\xi_P/L$ and $\xi_Q/L$ therefore coincide within statistical uncertainty. 
For $\Delta_Q>\Delta$, shown in the remaining panels, the two observables are no longer identical, but their finite-size crossing locations remain consistent with the same critical point. 
We therefore find a single FM--PM transition with no resolved intermediate spin-glass regime.

Figure~\ref{fig:splittransitionD} shows the complementary regime $\Delta_Q<\Delta$. 
For $\Delta/\Delta_Q=1.3$, the $\xi_P/L$ and $\xi_Q/L$ crossing locations separate, producing an intermediate overconfident phase. 
We also fix $\Delta=2$ and vary $\Delta_Q$. 
In this scan, $\xi_P/L$ remains noncritical throughout the parameter range, while $\xi_Q/L$ exhibits a crossing, realizing a direct PM--SG transition.

The transition estimates entering Fig.~\ref{fig:phase}(a) are obtained from these finite-size crossing sequences. 
The two limiting endpoints not obtained from the present scans are fixed by the zero-temperature three-dimensional Gaussian-RBIM transition~\cite{Krzakala2002Absence} and the critical coupling of the pure three-dimensional Ising model~\cite{Ferrenberg2018Pushing}.

\begin{figure}
    \centering
    \includegraphics[width=\columnwidth]{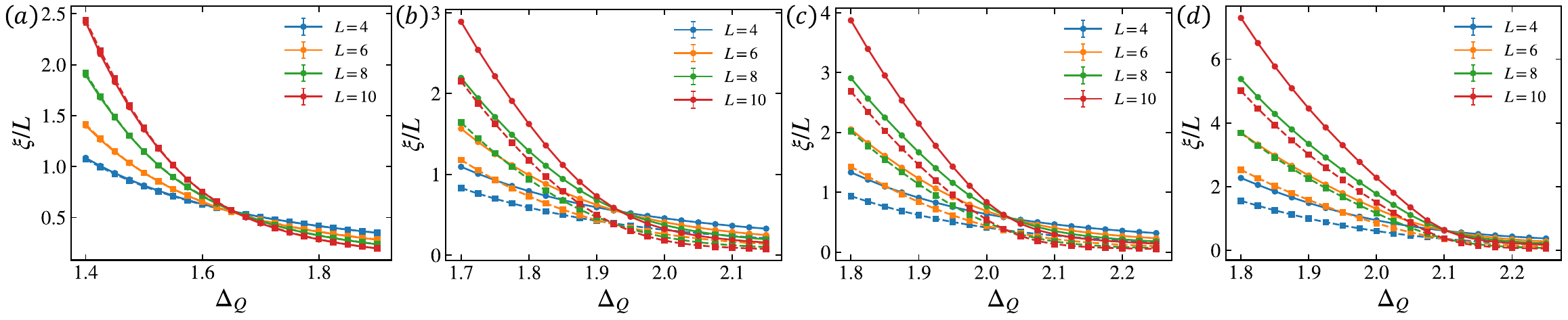}
    \caption{Finite-size crossings for noise mismatch in three dimensions.
    We set $\beta=\beta_Q=0$ and vary $\Delta_Q$ while keeping the ratio $\Delta/\Delta_Q$ fixed at (a) $1$, (b) $0.7$, (c) $0.5$, and (d) $0.25$. 
    Solid curves show $\xi_P/L$, associated with FM order ( truth--posterior reconstruction), while dashed curves show $\xi_Q/L$, associated with SG order (posterior--posterior overlap). 
    Different colors denote $L=4,6,8,10$, with $N=1792$ disorder realizations for each data point.
    The crossings determine the transition points used to construct the noise-mismatch phase diagram in Fig.~\ref{fig:phase}(a) in the main text.}
\label{fig:transitionD}
\end{figure}

\begin{figure}
    \centering
    \includegraphics[width=\columnwidth]{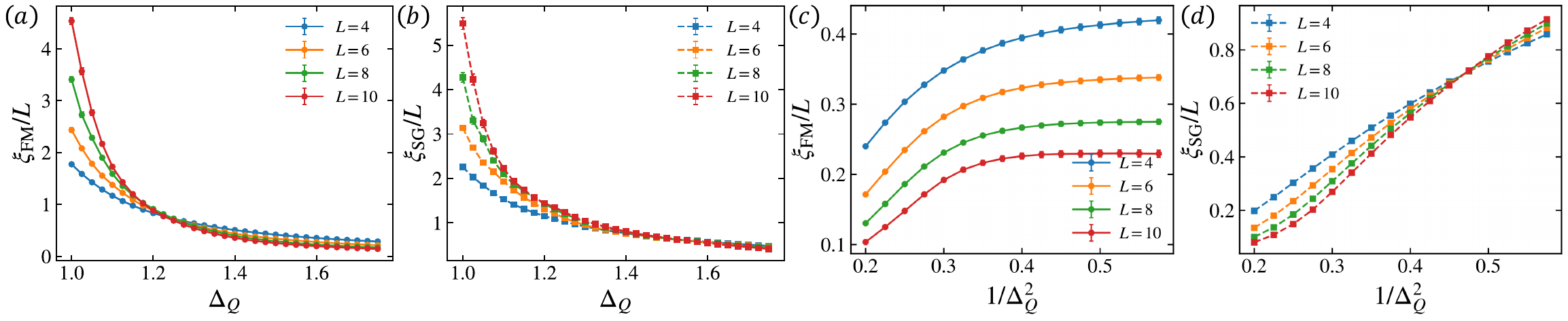}
    \caption{Separated FM and SG transitions for noise mismatch in three dimensions.
    (a,b) We set $\beta=\beta_Q=0$, fix $\Delta/\Delta_Q=1.3$, and vary $\Delta_Q$. 
    Panel (a) shows $\xi_P/L$, associated with FM order (truth--posterior reconstruction), while panel (b) shows $\xi_Q/L$, associated with SG order (posterior--posterior overlap).
    The two crossing locations are distinct, revealing an intermediate overconfident phase.
    (c,d) We fix $\Delta=2$ and vary $\Delta_Q$, using $\Delta_Q^{-2}$ as the tuning parameter.
    Panel (c) shows $\xi_P/L$ and remains in the PM regime throughout the scan, with no FM transition. 
    Panel (d) shows $\xi_Q/L$ and exhibits a SG transition.
    This scan therefore realizes a direct PM--SG transition.
    Different colors denote $L=4,6,8,10$, with $N=1792$ disorder realizations for each data point.} 
\label{fig:splittransitionD}
\end{figure}

\section{Two-dimensional results}
\label{sec:s_2d_numerics}

We next examine the same mismatched-inference problem in two dimensions.
Note that temperature mismatched in the two dimensional Ising model has been studied in Ref.~\cite{kim2026measurement}. 
In contrast to three dimensions, the two-dimensional Gaussian random-bond Ising model has no finite-temperature spin-glass phase. 
In the following, we use the two-dimensional case to test whether mismatch can produce a separated posterior-agreement transition at finite temperature. 
The relative-spin representation allows us to evaluate the required correlations using a free-Majorana transfer-matrix calculation.

\subsection{Free-Majorana calculation}
\label{sec:s_2d_majorana}

In two dimensions, we evaluate the relative-spin correlations using the free-Majorana representation of the Ising model rather than Monte Carlo sampling. 
The resulting two-dimensional random-bond Ising model is free fermionic, allowing its spin correlations to be evaluated using a Majorana transfer-matrix formulation.

For a row of $L$ spins, introduce $2L$ Majorana operators $\gamma_a$.
With an appropriate Jordan--Wigner convention,
\begin{equation}
    \sigma_x^x=i\gamma_{2x-1}\gamma_{2x},\qquad
    \sigma_x^z\sigma_{x+1}^z=i\gamma_{2x}\gamma_{2x+1}.
\label{eq:s_majorana_mapping}
\end{equation}
The free boundaries in the transfer direction select the even global spin-flip sector, in which the periodic horizontal bond is likewise represented by a Majorana bilinear. 
Both the horizontal Ising interactions and the vertical transfer operators are therefore Gaussian operators generated by Majorana bilinears.
The boundary state remains Gaussian under the row-by-row transfer and is completely characterized by its covariance matrix
\begin{equation}
    \Gamma_{ab}=\langle i\gamma_a\gamma_b\rangle.
\label{eq:s_majorana_covariance}
\end{equation}
The transfer matrix can consequently be implemented by updating $\Gamma$ rather than summing over the $2^L$ row configurations.

For an observable inserted at a given row, Gaussian states propagated from the two open boundaries define past and future environments with covariance matrices $\Gamma_{\rm P}$ and $\Gamma_{\rm F}$.
The normalized Majorana contractions between these environments are encoded in the transition covariance matrix
\begin{equation}
    \Gamma_{\rm tr}=i\left[(1-i\Gamma_{\rm F})(1-\Gamma_{\rm P}\Gamma_{\rm F})^{-1}(1-i\Gamma_{\rm P})-1\right].
\end{equation}
Equal-row Ising correlations are represented by Majorana strings, whose expectation values are reduced by the generalized Wick theorem to Pfaffians of the corresponding submatrices of $\Gamma_{\rm tr}$. 
In this way, $\langle\sigma_x\sigma_{x+r}\rangle$ is obtained exactly for every position and measurement realization.

For each fixed measurement record $\bm M$, we evaluate the relative-spin correlator $\langle\sigma_x\sigma_{x+r}\rangle_Q$. 
The two learning correlations are then obtained as
\begin{equation}
    G_P(r)=\mathbb E[\langle\sigma_x\sigma_{x+r}\rangle_Q],\qquad
    G_Q(r)=\mathbb E[\langle\sigma_x\sigma_{x+r}\rangle_Q^2],
\end{equation}
with an additional average over positions. 
This avoids posterior Monte Carlo sampling and eliminates thermalization errors in the two-dimensional calculations.

To characterize the finite-size behavior, we define the correlation ratios
\begin{equation}
    R_P=\frac{G_P(L/2)}{G_P(L/4)},\qquad
    R_Q=\frac{G_Q(L/2)}{G_Q(L/4)}.
\label{eq:s_2d_ratios}
\end{equation}
Here $R_P$ probes FM order (truth--posterior reconstruction), while $R_Q$ probes SG order (posterior--posterior agreement). 
These dimensionless ratios distinguish long-range from short-range correlations without requiring an order-parameter amplitude, and their finite-size crossings are used to locate the corresponding transitions.

\begin{figure}[t]
    \centering
    \includegraphics[width=\columnwidth]{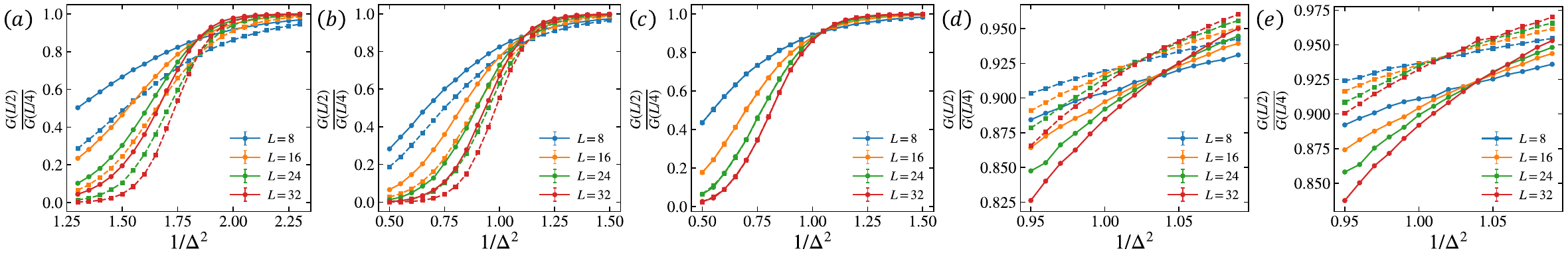}
    \caption{Finite-size transitions for noise mismatch in two dimensions.
    We set $\beta=\beta_Q=0$ and vary $\Delta$ while fixing $\Delta^2/\Delta_Q^2$ to (a) $0.3$, (b) $0.7$, (c) $1$, (d) $1.2$, and (e) $1.4$. Solid curves show the FM correlation ratio $G_P(L/2)/G_P(L/4)$, associated with truth--posterior reconstruction, while dashed curves show the SG correlation ratio $G_Q(L/2)/G_Q(L/4)$, associated with posterior--posterior overlap. Different colors denote $L=8,16,24,32$, with $N=8960$ disorder realizations for each data point.
    The two indicators are consistent with a single finite-temperature transition. The small apparent separation in panels (d) and (e) is attributed to stronger finite-size effects as the system approaches the zero-temperature SG regime of the two-dimensional Gaussian RBIM.}
\label{fig:transitionD_2D}
\end{figure}

\begin{figure}
    \centering
    \includegraphics[width=\columnwidth]{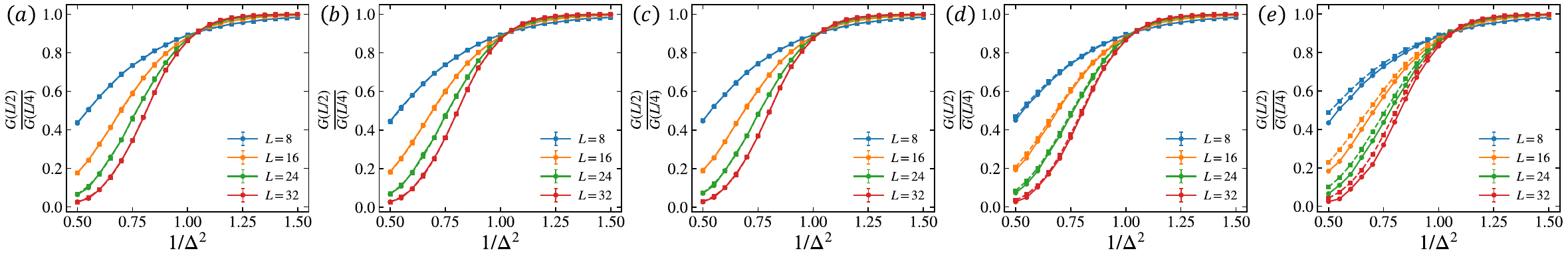}
    \caption{Finite-size crossings for temperature mismatch in two dimensions.
    We fix $\beta=0.1$ and matched noise $\Delta=\Delta_Q$, and vary $\Delta$ for (a) $\beta_Q=0$, (b) $\beta_Q=0.05$, (c) $\beta_Q=0.1$, (d) $\beta_Q=0.2$, and (e) $\beta_Q=0.3$.
    Solid curves show the FM correlation ratio $G_P(L/2)/G_P(L/4)$, associated with truth--posterior reconstruction, while dashed curves show the SG correlation ratio $G_Q(L/2)/G_Q(L/4)$, associated with posterior--posterior overlap. 
    Different colors denote $L=8,16,24,32$, with $N=8960$ disorder realizations for each data point.
    Across the parameter range studied, the FM and SG indicators do not develop distinct transition points, and only a direct FM--PM transition is observed. 
    This contrasts with the three-dimensional case, where a separate posterior-overlap transition produces an intermediate overconfident phase.
    }
\label{fig:transitionbeta_2D}
\end{figure}

\begin{figure}
    \centering
    \includegraphics[width=0.6\columnwidth]{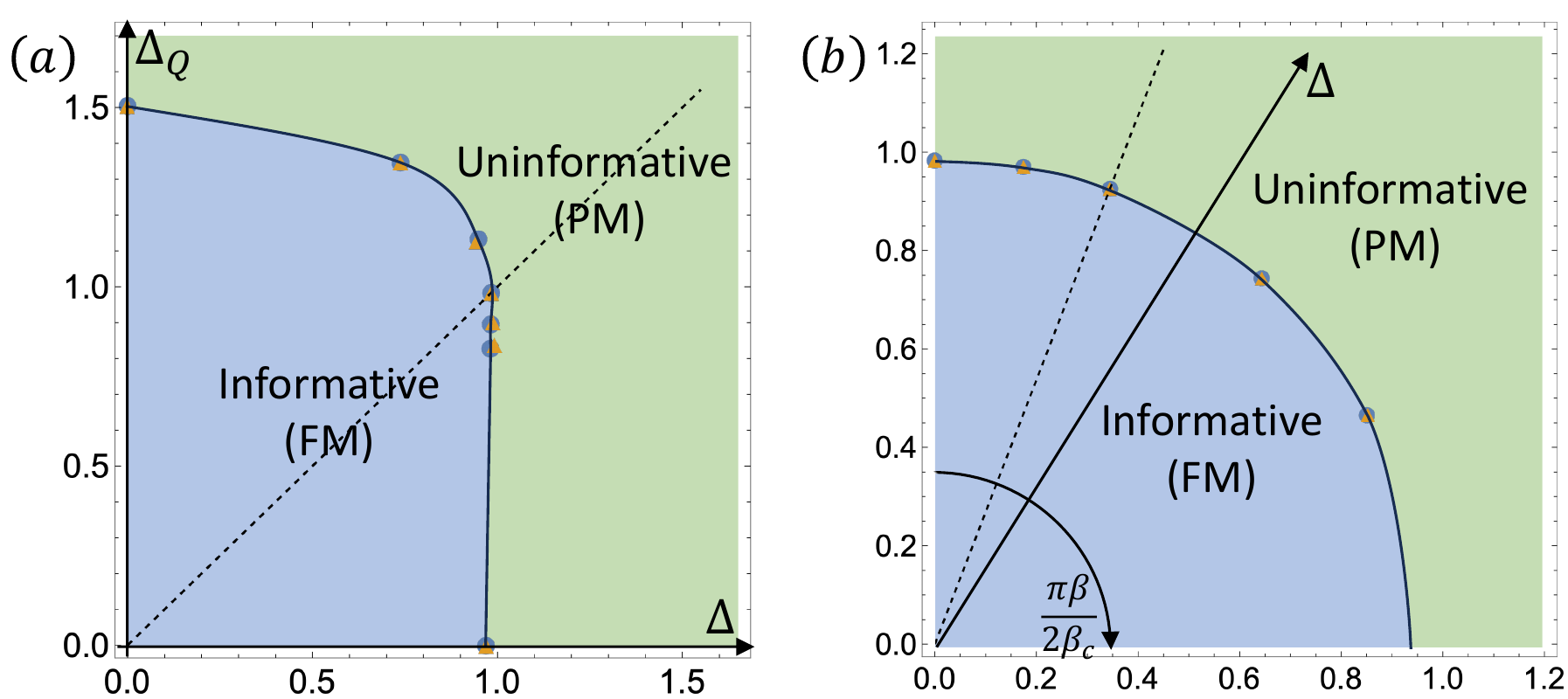}
    \caption{Numerical phase diagrams of mismatched Ising inference in two dimensions.
    (a) Noise mismatch at $\beta=\beta_Q=0$. We vary $\Delta$ while fixing $\Delta^2/\Delta_Q^2=0.3,0.7,1,1.2,1.4$. Blue circles and yellow triangles denote transition points extracted from the FM and SG indicators, respectively. The two sets of transition points coincide within numerical resolution, giving a single FM--PM boundary. The dashed line marks the Gaussian Nishimori condition $\Delta=\Delta_Q$.
    The transition point on the $\Delta_Q=0$ axis, obtained by tuning $\Delta$, is taken from Ref.~\cite{Melchert2013Information}.
    (b) Temperature mismatch at fixed $\beta=0.1$ and matched noise $\Delta=\Delta_Q$, for $\beta_Q=0,0.05,0.1,0.2,0.3$. As in the three-dimensional phase diagram, we use polar-like coordinates with radius $\Delta$ and angle $\theta=\pi\beta_Q/(2\beta_c)$, where $\beta_c=\ln(1+\sqrt{2})/2$ is the critical inverse temperature of the clean two-dimensional Ising model. The dashed ray corresponds to the Nishimori condition $\beta_Q=\beta$. Again, the FM and SG transition points coincide within numerical resolution. Thus, over the parameter ranges studied, only the informative (FM) and uninformative (PM) phases are observed in two dimensions, with no intermediate overconfident phase.}
\label{fig:phase_diagram_2D}
\end{figure}

\subsection{Two-dimensional crossings and phase diagrams}
\label{sec:s_2d_phase_diagram}

We summarize the numerical results for the two-dimensional model.
In contrast to the three-dimensional case, we do not find a resolved separation between the FM and SG transitions within the parameter ranges studied. 
Instead, the two indicators track each other closely and give a single transition line within numerical resolution.
This is consistent with the absence of a finite-temperature SG phase in the two-dimensional Gaussian random-bond Ising model.

Figure~\ref{fig:transitionD_2D} shows the noise-mismatch scans at $\beta=\beta_Q=0$. 
We vary $\Delta$ while fixing $\Delta^2/\Delta_Q^2= 0.3,0.7,1,1.2,1.4$. 
For the first three cases, the FM and SG order indicators are consistent with a single transition. 
In particular, for $\Delta^2/\Delta_Q^2=1$, corresponding to the Nishimori condition $\Delta=\Delta_Q$, the FM and SG curves coincide, as required by the Nishimori identity. 
A small separation develops in panels (d) and (e), where $\Delta^2/\Delta_Q^2>1$. 
We attribute this to finite-size effects.
In this regime the effective randomness is stronger and the system moves closer to the zero-temperature SG regime of the two-dimensional Gaussian RBIM, where finite-size corrections become more pronounced. 
The data do not show evidence for a stable intermediate SG phase at finite temperature.

Figure~\ref{fig:transitionbeta_2D} shows the matched-noise temperature-mismatch scans at fixed $\beta=0.1$ and $\Delta=\Delta_Q$, for $\beta_Q=0,0.05,0.1,0.2,0.3$. 
Again, the FM and SG indicators do not develop distinct crossings.
In particular, for $\beta_Q\leq\beta$, the two indicators nearly coincide over the parameter range shown.
Thus, unlike in three dimensions, temperature mismatch does not generate a numerically resolved overconfident regime in two dimensions.

The corresponding phase diagrams are summarized in Fig.~\ref{fig:phase_diagram_2D}. 
In both the noise-mismatch and temperature-mismatch cuts, the transition points extracted from the two channels coincide within numerical resolution. 
We therefore observe only the informative and uninformative phases in two dimensions over the parameter ranges studied.

\section{Mismatched inference in deformed toric codes}

Our mismatched-inference framework can be extended directly to the deformed toric-code measurement problem studied in Refs.~\cite{putz2026learning,patil2026higher}.
We first briefly review the matched construction and then allow the parameters used to interpret the measurement outcomes to differ from those that generate the physical state and measurement record.

We consider the toric code on a square lattice, with a qubit on each link $\langle ij\rangle$. 
We use the convention

\begin{equation}
    H_{\rm TC}=-\sum_v A_v-\sum_p B_p,\qquad A_v=\prod_{\ell\ni v}\hat X_\ell,\qquad B_p=\prod_{\ell\in\partial p}\hat Z_\ell , 
\end{equation}
and denote by $|{\rm TC}\rangle$ a ground state.
We write $\hat Z_{ij}$ for the Pauli-$Z$ operator on the link connecting sites $i$ and $j$. 
With this convention, the factor $\beta/2$ in the wave-function amplitude is chosen such that its norm carries the classical Ising weight $e^{\beta\sum_{\langle ij\rangle}\sigma_i\sigma_j}$.
Consider the deformed toric-code wavefunction 
\begin{equation}
    |\mathrm{TC}\rangle_\beta
    =\frac{1}{\sqrt{\mathcal N_\beta}}
    \exp\left(\frac{\beta}{2}\sum_{\langle ij\rangle}\hat Z_{ij}\right)
    |\mathrm{TC}\rangle ,
\label{eq:dtc_state}
\end{equation}
Weak measurements of $\hat Z_{ij}$ are described by the Kraus operators~\cite{putz2026learning,patil2026higher}
\begin{equation}
    \hat K_{\bm m}(\gamma)
    =\frac{\exp\left[\frac{\widetilde\gamma}{2}\sum_{\langle ij\rangle}m_{ij}\hat Z_{ij}\right]}
    {(2\cosh\widetilde\gamma)^{N_b/2}},
    \qquad m_{ij}=\pm1,\qquad \gamma=\tanh\widetilde\gamma,
\label{eq:dtc_Kraus}
\end{equation}
where $\bm m=\{m_{ij}\}$ is the measurement record and $N_b$ is the number of links. 
The physical Born probability is
\begin{equation}
    P_{\beta,\gamma}(\bm m)
    ={}_\beta\langle\mathrm{TC}|
    \hat K_{\bm m}^\dagger(\gamma)\hat K_{\bm m}(\gamma)
    |\mathrm{TC}\rangle_\beta,
\label{eq:dtc_Born}
\end{equation}
and the corresponding normalized postmeasurement state is
\begin{equation}
    |\Psi^P_{\bm m}\rangle
    =\frac{\hat K_{\bm m}(\gamma)|\mathrm{TC}\rangle_\beta}
    {\sqrt{P_{\beta,\gamma}(\bm m)}}.
\label{eq:dtc_true_state}
\end{equation}

To introduce mismatch, we keep the physical record $\bm m$ generated with the true parameters $(\beta,\gamma)$, but interpret the same record using independent inferred parameters $(\beta_Q,\gamma_Q)$. 
We define
\begin{equation}
    |\Psi^Q_{\bm m}\rangle
    =\frac{\hat K_{\bm m}(\gamma_Q)|\mathrm{TC}\rangle_{\beta_Q}}
    {\sqrt{\mathcal N_Q(\bm m)}},
\label{eq:dtc_inferred_state}
\end{equation}
where
\begin{equation}
    \mathcal N_Q(\bm m)
    ={}_{\beta_Q}\langle\mathrm{TC}|
    \hat K_{\bm m}^\dagger(\gamma_Q)\hat K_{\bm m}(\gamma_Q)
    |\mathrm{TC}\rangle_{\beta_Q},
    \qquad \gamma_Q=\tanh\widetilde\gamma_Q.
\end{equation}
The state $|\Psi^Q_{\bm m}\rangle$ is not the physical postmeasurement state unless $(\beta_Q,\gamma_Q)=(\beta,\gamma)$.
Rather, it represents the state reconstructed from the observed trajectory using a mismatched model.

For a fixed measurement trajectory, we define the true and inferred conditional expectation values by
\begin{equation}
    \langle\hat{\mathcal O}\rangle_{P,\bm m}
    \equiv\langle\Psi^P_{\bm m}|\hat{\mathcal O}|\Psi^P_{\bm m}\rangle,\qquad
    \langle\hat{\mathcal O}\rangle_{Q,\bm m}
    \equiv\langle\Psi^Q_{\bm m}|\hat{\mathcal O}|\Psi^Q_{\bm m}\rangle.
\label{eq:dtc_conditional_expectation}
\end{equation}
We use an overbar to denote an average over the \emph{physical} measurement trajectories,
\begin{equation}
    \overline{F(\bm m)}
    \equiv\sum_{\bm m}P_{\beta,\gamma}(\bm m)F(\bm m).
\label{eq:dtc_bar}
\end{equation}
Thus the overbar is always evaluated using the true Born distribution determined by $(\beta,\gamma)$, while the subscripts $P$ and $Q$ specify whether the conditional expectation value is evaluated using the physical or inferred state.

With this notation, the measurement-averaged moments considered in Ref.~\cite{patil2026higher} generalize directly to
\begin{equation}
    \overline{\langle\hat{\mathcal O}_1\rangle_Q\cdots
    \langle\hat{\mathcal O}_N\rangle_Q}
    =\sum_{\bm m}P_{\beta,\gamma}(\bm m)
    \prod_{r=1}^N\langle\hat{\mathcal O}_r\rangle_{Q,\bm m}.
\label{eq:dtc_Q_moment}
\end{equation}
We can similarly define mixed truth--inference moments,
\begin{equation}
    \overline{\langle\hat{\mathcal O}_0\rangle_P
    \langle\hat{\mathcal O}_1\rangle_Q\cdots
    \langle\hat{\mathcal O}_N\rangle_Q}
    =\sum_{\bm m}P_{\beta,\gamma}(\bm m)
    \langle\hat{\mathcal O}_0\rangle_{P,\bm m}
    \prod_{r=1}^N\langle\hat{\mathcal O}_r\rangle_{Q,\bm m}.
\label{eq:dtc_PQ_moment}
\end{equation}
In particular, the analogues of the truth--posterior and posterior--posterior correlations introduced above are $C_P^{\mathcal O}   =\overline{\langle\hat{\mathcal O}\rangle_P\langle\hat{\mathcal O}\rangle_Q}$, $ C_Q^{\mathcal O}=\overline{\langle\hat{\mathcal O}\rangle_Q^2}$.
At matching $\beta_Q=\beta,\ \gamma_Q=\gamma$, the two conditional states coincide trajectory by trajectory, so $\langle\hat{\mathcal O}\rangle_{P,\bm m}=\langle\hat{\mathcal O}\rangle_{Q,\bm m}$ and $C_P^{\mathcal O}=C_Q^{\mathcal O}$.

The connection to our classical mismatched-inference formulation is most transparent in the dual Ising representation.
The dual Ising description follows by resolving the toric-code configurations in the $\hat Z$ basis. Within the flux-free sector, the link variables can be parameterized as
$Z_{ij}=\sigma_i\sigma_j$,
where $\sigma_i=\pm1$ are Ising spins living on the sites of the dual square lattice.
The true prior is
\begin{equation}
    P_\beta(\bm\sigma)
    =\frac{1}{Z_\beta}e^{-\beta H(\bm\sigma)},\qquad
    Z_\beta=\sum_{\bm\sigma}e^{-\beta H(\bm\sigma)}, \qquad H(\bm\sigma)=-\sum_{\langle ij\rangle}\sigma_i\sigma_j.
\label{eq:dtc_true_prior}
\end{equation}
For a given configuration, the binary measurement outcomes are independently distributed according to
\begin{equation}
    P_\gamma(\bm m|\bm\sigma)
    =\prod_{\langle ij\rangle}
    \frac{1+\gamma m_{ij}\sigma_i\sigma_j}{2}.
\label{eq:dtc_true_channel}
\end{equation}
The physical distribution of measurement records is therefore
\begin{equation}
    P_{\beta,\gamma}(\bm m)
    =\sum_{\bm\sigma}P_\beta(\bm\sigma)
    P_\gamma(\bm m|\bm\sigma).
\label{eq:dtc_record_probability}
\end{equation}

Conditioned on $\bm m$, the true posterior is
\begin{equation}
    P_{\beta,\gamma}(\bm\sigma|\bm m)
    =\frac{1}{Z_P(\bm m)}
    \exp\left[-\beta H(\bm\sigma)
    +\widetilde\gamma\sum_{\langle ij\rangle}
    m_{ij}\sigma_i\sigma_j\right],
    \qquad
    Z_P(\bm m)
    =\sum_{\bm\sigma}
    \exp\left[-\beta H(\bm\sigma)
    +\widetilde\gamma\sum_{\langle ij\rangle}
    m_{ij}\sigma_i\sigma_j\right].
\label{eq:dtc_true_posterior}
\end{equation}
The corresponding mismatched posterior is
\begin{equation}
    Q_{\beta_Q,\gamma_Q}(\bm\sigma|\bm m)
    =\frac{1}{Z_Q(\bm m)}
    \exp\left[-\beta_QH(\bm\sigma)
    +\widetilde\gamma_Q\sum_{\langle ij\rangle}
    m_{ij}\sigma_i\sigma_j\right], \qquad 
    Z_Q(\bm m)
    =\sum_{\bm\sigma}
    \exp\left[-\beta_QH(\bm\sigma)
    +\widetilde\gamma_Q\sum_{\langle ij\rangle}
    m_{ij}\sigma_i\sigma_j\right].
\label{eq:dtc_mismatched_posterior}
\end{equation}
Thus $\beta_Q\neq\beta$ corresponds to using an incorrect prior, or equivalently an incorrect deformation parameter in the quantum problem, while $\gamma_Q\neq\gamma$ corresponds to using an incorrectly calibrated measurement strength.

To study the inference for the logical information, we first consider the trajectory-resolved domain-wall entropy of the deformed toric code~\cite{putz2026learning,patil2026higher}. 
For a measurement outcome $\bm m$, let $Z_{\bm m}^{(+)}$ and $Z_{\bm m}^{(-)}$ denote the partition functions associated with the two logical states $\left|{{\rm TC},+}\right>_\beta$ and $\left|{{\rm TC},-}\right>_\beta$ on a cylinder. 
The conditional entropy of the logical sector is
\begin{equation}
    I_{\bm m}=-p_+(\bm m)\log_2 p_+(\bm m)-p_-(\bm m)\log_2 p_-(\bm m),
\end{equation}
where
\begin{equation}
    p_\pm(\bm m)=\frac{Z_{\bm m}^{(\pm)}}{Z_{\bm m}^{(+)}+Z_{\bm m}^{(-)}}.
\end{equation}
Its measurement average $I_c=\overline{I_{\bm m}}$ is equivalent to the coherent information considered in Ref.~\cite{putz2026learning}.

Under the mapping to the classical Ising model, the two logical sectors correspond to the two boundary conditions $\sigma_L\sigma_R=\pm1$, where auxiliary boundary spins $\sigma_L$ and $\sigma_R$ are coupled to all spins on the corresponding open boundary. 
Then, one defines the boundary-to-boundary correlation
\begin{equation}
    C_{\bm m}\equiv\langle\sigma_L\sigma_R\rangle_{\bm m}=\frac{Z_{\bm m}^{(+)}-Z_{\bm m}^{(-)}}{Z_{\bm m}^{(+)}+Z_{\bm m}^{(-)}}.
\end{equation}
It follows that
\begin{equation}
    I_{\bm m}=-\frac{1+C_{\bm m}}{2}\log_2\frac{1+C_{\bm m}}{2}-\frac{1-C_{\bm m}}{2}\log_2\frac{1-C_{\bm m}}{2}.
\end{equation}
Thus the boundary correlation $C_{\bm m}$ directly probes the distinguishability of the two logical sectors~\cite{putz2026learning}.

The boundary correlator provides the direct logical-information diagnostic. 
To characterize the corresponding ordering transition in the thermodynamic limit, it is useful to consider the same conditional correlation for two generic, widely separated spins, $\langle\sigma_i\sigma_j\rangle_{\bm m}$.
Long-range order of these bulk correlations diagnoses the  PM, ``spin-glass'', and FM phases that control the scaling of the boundary logical correlator~\cite{putz2026learning,patil2026higher}. 
Note that the terminologies for the phases in the corresponding statistical model are different in our paper, as will be clarified in the following. 
This formulation also allows a natural extension to mismatched inference. 
We define
\begin{equation}
    C_{P}(i,j)=\overline{\langle\sigma_i\sigma_j \rangle_{P}\langle\sigma_i\sigma_j\rangle_{Q}},
    \qquad C_{Q}(i,j)=\overline{\langle\sigma_i \sigma_j\rangle_{Q}^{2}},
\label{eq:suboptimal correlator}
\end{equation}
where $\langle O\rangle_P$ and $\langle O\rangle_Q$ denote conditional averages for the same measurement outcome $\bm m$, evaluated using the true parameters $(\beta,\gamma)$ and the assumed parameters $(\beta_Q,\gamma_Q)$, respectively.
The overline denotes an average over $\bm m$ drawn from the true measurement distribution.
For the boundary choice $(i,j)=(L,R)$, these quantities directly compare the logical-sector information inferred by the true and mismatched conditional states. 
For generic bulk sites, they serve as long-distance order parameters for the corresponding learning phases.

The cross-correlation $C_P$ measures whether the correlations inferred by the mismatched model agree with those of the true conditional state, while $C_Q$ measures the internal confidence of the mismatched model irrespective of whether that inference is correct.
At Bayes matching, $(\beta_Q,\gamma_Q)=(\beta,\gamma)$, the two conditional distributions coincide trajectory by trajectory, so that
\begin{equation}
    C_P(i,j)=C_Q(i,j)=\overline{\langle\sigma_i\sigma_j\rangle_P^2}.
\end{equation}
This recovers the Edwards--Anderson nonlinear correlation used to diagnose the learning transition in Refs.~\cite{putz2026learning,patil2026higher}.

Away from Bayes matching, $C_P$ and $C_Q$ can separate.
In the thermodynamic limit, $C_P=C_Q=0$ defines an uninformative (toric code) phase. 
A phase with $C_P=0$ but $C_Q>0$ is logically overconfident.
The assumed model develops long-range conditional order, and hence a definite trajectory-dependent logical-sector bias in the boundary formulation, without correctly tracking the true conditional state. 
Finally, $C_P>0$, together with $C_Q>0$, corresponds to an informative (dephased) phase in which the inferred long-range order remains correlated with the physical logical information.

For the classical model, the general measurement-averaged moments of operators can be expressed as
\begin{equation}
    \overline{\langle O_0\rangle_P\langle O_1\rangle_Q\cdots\langle O_N\rangle_Q}
    =\sum_{\bm \sigma^0,\bm m} P_{\beta,\gamma}(\bm m)
    P_{\beta,\gamma}(\bm\sigma^0|\bm m)\sum_{\bm \sigma^1,\ldots,\bm \sigma^N} O_0\otimes\cdots O_N\prod_{\alpha=1}^N Q_{\beta_Q,\gamma_Q}(\bm\sigma^\alpha|\bm m).
\label{eq:sigma_posterior_moment}
\end{equation}
Similar to the case above, introducing $n$ replicas with the limit $n\rightarrow0$, we have
\begin{equation}
\begin{split}
    \overline{\langle O_0\rangle_P\langle O_1\rangle_Q\cdots\langle O_N\rangle_Q}
    =\lim_{n\to0}\frac{(2\cosh{\Tilde{\gamma}})^{-N_b}}{Z_\beta}\sum_{\bm m}\sum_{\bm \sigma^0,\bm \sigma^1,\ldots,\bm \sigma^n} & e^{-\beta H(\bm\sigma^0)}e^{\Tilde{\gamma}\sum_{\langle ij\rangle}m_{ij}\sigma^0_i\sigma^0_j}\,
    O_0\otimes\cdots\otimes O_N\\
    &\times\prod_{\alpha=1}^n 
    e^{-\beta_QH(\bm\sigma^\alpha)
    +\widetilde\gamma_Q\sum_{\langle ij\rangle}
    m_{ij}\sigma_i^\alpha\sigma_j^\alpha},
\label{eq:s_replica_identity_tc}
\end{split}
\end{equation}
where $N_b$ is the number of bonds.
Replica $0$ denotes the physical true configuration and is never analytically continued, and the number of other replicas is analytically continued to zero.

This form makes the role of mismatch transparent.
The replicated weight before averaging over the measurement record is, up to normalization factors independent of the spins,
\begin{equation}
    \exp\left[-\beta H(\bm\sigma^0)-\beta_Q\sum_{\alpha=1}^n H(\bm\sigma^\alpha)+\sum_{\langle ij\rangle}m_{ij}
    \left(\widetilde\gamma\,\sigma_i^0\sigma_j^0 +\widetilde\gamma_Q\sum_{\alpha=1}^n\sigma_i^\alpha\sigma_j^\alpha\right)\right].
\label{eq:dtc_replica_weight_m}
\end{equation}
Defining the bond variables $x_{ij}^a=\sigma_i^a\sigma_j^a$, the sum over each binary measurement outcome can be carried out exactly.
The exact replicated lattice action is therefore
\begin{equation}
    -\mathcal H_n=\beta\sum_{\langle ij\rangle}x_{ij}^0+\beta_Q\sum_{\alpha=1}^n\sum_{\langle ij\rangle}x_{ij}^\alpha+\sum_{\langle ij\rangle}
    \ln\left[2\cosh\left(\widetilde\gamma x_{ij}^0+\widetilde\gamma_Q\sum_{\alpha=1}^n x_{ij}^\alpha\right)\right],
\label{eq:dtc_replica_action}
\end{equation}
up to an additive constant.


For weak measurements, Eq.~\eqref{eq:dtc_replica_action} can be expanded with small $\Tilde{\gamma}$ and $\Tilde{\gamma}_Q$.
The leading measurement-induced interactions are
\begin{equation}
    \lambda_P\sum_{\alpha=1}^n\sum_{\langle ij\rangle}x_{ij}^0x_{ij}^\alpha
    +\lambda_Q\sum_{\alpha<\gamma}\sum_{\langle ij\rangle}x_{ij}^\alpha x_{ij}^\gamma+\cdots ,
    \qquad \lambda_P=\widetilde\gamma\,\widetilde\gamma_Q,\qquad
    \lambda_Q=\widetilde\gamma_Q^2.
\label{eq:dtc_replica_action_weak}
\end{equation}
The pairwise terms in Eq.~\eqref{eq:dtc_replica_action_weak} reproduce precisely the same replica structure as the Gaussian mismatched-inference theory considered above.
In particular, at matching, $\lambda_P=\lambda_Q=\widetilde\gamma^2$, so the truth--inference and inference--inference overlap sectors become equivalent.
A measurement-strength mismatch instead produces $\frac{\lambda_Q}{\lambda_P}=\frac{\widetilde\gamma_Q}{\widetilde\gamma}$, while a deformation mismatch $\beta\neq\beta_Q$ independently splits the single-replica couplings. 
The deformed toric-code problem therefore realizes the same general structure as our classical mismatched-learning theory. 


To make the phase interpretation transparent, it is useful to perform the same relative-spin change of variables as in the classical inference problem. 
Define
$\tau_i=\sigma_i^0\sigma_i$, $ \eta_{ij}=m_{ij}\sigma_i^0\sigma_j^0$. 
The transformed measurement variables are independently distributed as
\begin{equation}
    P_\gamma(\eta_{ij}) =\frac{1+\gamma\eta_{ij}}{2} =\frac{e^{\widetilde\gamma\eta_{ij}}}{2\cosh\widetilde\gamma}, \qquad \eta_{ij}=\pm1, 
\end{equation}
Using $\sigma_i\sigma_j =\sigma_i^0\sigma_j^0\,\tau_i\tau_j$, $m_{ij}\sigma_i\sigma_j =\eta_{ij}\tau_i\tau_j$, the mismatched posterior becomes a random-bond Ising model for the relative spins,
\begin{equation}            
    Q(\tau|\sigma^0,\eta) = \frac{1}{Z_Q} \exp\!\left[ \sum_{\langle ij\rangle} J_{ij}\tau_i\tau_j \right], \qquad J_{ij} = \beta_Q\sigma_i^0\sigma_j^0 +\widetilde\gamma_Q\eta_{ij}
\label{eq:transform_TC}
\end{equation}
where quenched disorder is generated by the true Ising configuration $\sigma^0\sim P_\beta$ together with the independent binary variables $\eta_{ij}$. 
In the special case $\beta=\beta_Q=0$, Eq.~\eqref{eq:transform_TC} reduces to the standard two-dimensional $\pm J$ random-bond Ising model.

The correlators in Eq.~\eqref{eq:suboptimal correlator} acquire a particularly simple interpretation in this relative-spin representation. 
Using Bayes' rule for the true conditional average,
\begin{equation} 
    C_P(i,j) = \overline{ \langle\sigma_i\sigma_j\rangle_P \langle\sigma_i\sigma_j\rangle_Q } = \mathbb E_{\sigma^0,\eta} \!\left[ \langle\tau_i\tau_j\rangle_J \right], \qquad 
    C_Q(i,j) = \overline{ \langle\sigma_i\sigma_j\rangle_Q^2 } = \mathbb E_{\sigma^0,\eta} \!\left[ \langle\tau_i\tau_j\rangle_J^2 \right]. 
\end{equation}
Thus, $C_P$ is the ferromagnetic correlation of the transformed model, whereas $C_Q$ is its two-replica spin-glass correlation. 
The three possible long-distance behaviors therefore have the same interpretation as in the classical mismatched-inference problem: the paramagnetic (PM) phase, $C_P=C_Q=0$, corresponds to an unlearned logical degree of freedom; the ferromagnetic (FM) phase, $C_P>0$ and $C_Q>0$, corresponds to correctly learned logical information; and a spin-glass (SG) phase with $C_P=0$ but $C_Q>0$ would correspond to logical overconfidence.

It is useful to clarify the relation between the present mismatch classification and the matched learning phases studied in Refs.~\cite{putz2026learning,patil2026higher}.
In the matched problem, the phases are classified using the original dual-Ising spins $\sigma$. 
The paramagnetic, ``spin-glass'', and ferromagnetic regimes correspond respectively to the toric code, dephased, and trivial phases of the measured deformed toric code.  
In particular, the ``spin-glass'' phase in this terminology is characterized by vanishing ordinary magnetic order but nonzero nonlinear order,
\begin{equation}
    \overline{\langle \sigma_i\sigma_j\rangle_P}=0,
    \qquad
    \overline{\langle \sigma_i\sigma_j\rangle_P^2}\neq 0
\end{equation}
at long distances.

Our mismatch diagnostics characterize a different question. 
Here, we consider the phase diagram of the mismatched inference problem below the threshold in the deformed toric code, or equivalently, $\beta, \beta_Q < \beta_c$ in the dual Ising model. 
At Bayes matching, the physical and inferred conditional distributions coincide, so
\begin{equation}
    C_P(i,j)=C_Q(i,j)=
    \overline{\langle \sigma_i\sigma_j\rangle_P^2}.
\end{equation}
Consequently, $C_P$ and $C_Q$ alone do not distinguish the ``spin-glass'' (dephased) and ferromagnetic (trivial) phases of the matched problem, since both have nonzero nonlinear order. 
Their distinction additionally requires the ordinary dual-Ising correlation $\overline{\langle \sigma_i\sigma_j\rangle_P}$, which is trivial for $\beta, \beta_Q < \beta_c$. 
After the relative-spin transformation $\tau_i=\sigma_i^0\sigma_i$, the matched nonlinear correlation becomes ferromagnetic reconstruction order of the relative spins. 
By contrast, the overconfident phase considered here is characterized by
\begin{equation}
    C_P=0, \qquad C_Q>0,
\end{equation}
and therefore corresponds to genuine spin-glass order of the relative-spin model without ferromagnetic reconstruction order.
Thus the ``spin-glass'' (dephased) phase of the matched learning problem~\cite{putz2026learning,patil2026higher} and the overconfident spin-glass phase discussed here refer to distinct order parameters and should not be identified.

Finally, the transformed representation explains why a stable logically overconfident phase is not expected at finite effective temperature in two dimensions. 
Such a phase would require SG order of the short-range two-dimensional random-bond Ising model without FM order. 
In the $\beta=\beta_Q=0$ limit this is precisely the conventional two-dimensional $\pm J$ model, whose SG transition occurs only at zero temperature. 
Thus, in the present mismatch problem, overconfident SG order can occur only in the zero-effective-temperature limit, corresponding to $\widetilde{\gamma}_Q\rightarrow\infty$ or $\gamma_Q\rightarrow1$.
This SG order should not be confused with the ``spin-glass'' (dephased) phase of the matched learning problem in Refs.~\cite{putz2026learning,patil2026higher}, which is defined by nonlinear order of the original dual-Ising variables rather than by SG order of the relative spins, as we clarified before.
More generally, for $\beta<\beta_c$ the disorder inherited from the planted Ising configuration is short-range correlated, so a stable finite-temperature SG phase is likewise not expected. 

\section{Learning the unknown parameters}
\label{sec:s_parameter_learning}

For the Gaussian channel considered here, the first two record moments already determine $\beta$ and $\Delta$. 
We then present evidence maximization as a more general parameter-learning procedure.

\subsection{Direct moment determination for the Gaussian channel}
\label{sec:s_moment_learning}

For each bond, $M_{ij}=s_i^0s_j^0+\Delta z_{ij}$ with $z_{ij}\sim\mathcal N(0,1)$. 
Define the bond-averaged moments $m_k=N_b^{-1}\sum_{\langle ij\rangle}\E[M_{ij}^k]$ and the clean-Ising nearest-neighbor correlation $e(\beta)=N_b^{-1}\langle\sum_{\langle ij\rangle}s_is_j\rangle_\beta$.
Since the Gaussian noise is centered and independent of the true spins, while $(s_i^0s_j^0)^2=1$, one obtains
\begin{equation}
    m_1=e(\beta),\qquad m_2=1+\Delta^2.
\label{eq:s_moment_result}
\end{equation}
Thus the second moment directly determines the noise strength, $\Delta^2=m_2-1$, while the first moment determines $\beta$ through the clean-Ising equation of state.

To see that this inversion is unique, define $E(\bm s)=\sum_{\langle ij\rangle}s_is_j$. 
Differentiating the clean-Ising partition function gives
\begin{equation}
    \frac{\dd e}{\dd\beta}=\frac1{N_b}\operatorname{Var}_\beta[E]\ge0.
\label{eq:s_energy_monotonicity}
\end{equation}
For a finite nontrivial system, the variance is strictly positive, so $e(\beta)$ is strictly increasing and uniquely determines $\beta$. 
The first two record moments therefore determine $\beta$ and $\Delta$ separately without requiring posterior inference.

For $R_m$ independent measurement records, define the empirical moments
\begin{equation}
    \widehat m_k=\frac1{R_m N_b}\sum_{r=1}^{R_m}\sum_{\langle ij\rangle}\left(M_{ij}^{(r)}\right)^k.
\end{equation}
The corresponding estimators are $\widehat\Delta^2=\widehat m_2-1$ and $\widehat\beta=e^{-1}(\widehat m_1)$. 
As $R_m\to\infty$, the empirical moments converge to Eq.~\eqref{eq:s_moment_result}, establishing consistency of these estimators.

\subsection{Evidence maximization}
\label{sec:s_evidence_learning}

The moment relations above exploit special properties of the binary signal and additive Gaussian channel.
A more general approach is to determine the assumed parameters by maximizing the probability of the observed measurement records. 
Summing Eq.~\eqref{eq:s_assumed_joint_expanded} over the latent spins gives the record evidence
\begin{equation}
    p_Q(\bm M)=\frac{Z_Q(\bm M)}{Z_{\beta_Q}(2\pi\Delta_Q^2)^{N_b/2}}
    \exp\!\left[-\frac1{2\Delta_Q^2}\sum_b(M_b^2+1)\right].
\label{eq:s_evidence}
\end{equation}
If $p_{\beta,\Delta}(\bm M)$ denotes the true distribution of measurement records, the expected log evidence of the assumed model is
\begin{equation}
    \E_M[\ln p_Q(\bm M)]=\int\dd\bm M\,p_{\beta,\Delta}(\bm M)\ln p_Q(\bm M).
\label{eq:s_population_evidence}
\end{equation}
The expected log evidence is therefore maximized with respect to the assumed parameters $(\beta_Q,\Delta_Q)$.
Using the definition of the Kullback--Leibler (KL) divergence,
\begin{equation}
    D_{\rm KL}\!\left[p_{\beta,\Delta}\Vert p_Q\right]=\int\dd\bm M\,p_{\beta,\Delta}(\bm M)\ln\frac{p_{\beta,\Delta}(\bm M)}{p_Q(\bm M)}=-H[p_{\beta,\Delta}]-\E_M[\ln p_Q(\bm M)],
\end{equation}
where $H[p_{\beta,\Delta}]=-\int\dd\bm M\,p_{\beta,\Delta}(\bm M)\ln p_{\beta,\Delta}(\bm M)$ is the entropy of the true record distribution. 
Therefore
\begin{equation}
    \E_M[\ln p_Q(\bm M)]=-D_{\rm KL}\!\left[p_{\beta,\Delta}\Vert p_Q\right]-H[p_{\beta,\Delta}].
\label{eq:s_KL_evidence}
\end{equation}
Since $H[p_{\beta,\Delta}]$ is independent of $(\beta_Q,\Delta_Q)$, maximizing the expected log evidence is equivalent to minimizing the KL divergence between the true and assumed record distributions. 
Because the assumed model contains the true distribution, the KL divergence vanishes at $(\beta_Q,\Delta_Q)=(\beta,\Delta)$.

To determine the optimal assumed parameters, we set the derivatives of the expected log evidence to zero. 
For a fixed measurement record $\bm M$, the corresponding derivatives of the log evidence are
\begin{equation}
    \frac{\partial\ln p_Q(\bm M)}{\partial\beta_Q}
    =\langle E\rangle_Q-\langle E\rangle_{\beta_Q},\qquad
    \frac{\partial\ln p_Q(\bm M)}{\partial\Delta_Q}
    =-\frac{N_b}{\Delta_Q}+\frac1{\Delta_Q^3}\left\langle\sum_b(M_b-s_is_j)^2\right\rangle_Q,
\label{eq:s_scores}
\end{equation}
where $E(\bm s)=\sum_b s_is_j$.
Here $\langle\cdots\rangle_Q$ denotes the posterior expectation at fixed record $\bm M$ under the assumed parameters $(\beta_Q,\Delta_Q)$, while $\langle\cdots\rangle_{\beta_Q}$ denotes the thermal expectation in the clean Ising model at inverse temperature $\beta_Q$.
The first equation matches the bond energy inferred from the record to that of the assumed clean-Ising prior, while the second matches the posterior-averaged channel residual to the assumed noise variance.

For $R_m$ independent records $\bm M^{(r)}$, maximizing the empirical average log evidence requires
\begin{equation}
    \frac1{R_m}\sum_{r=1}^{R_m}\langle E\rangle_{Q_r}=\langle E\rangle_{\beta_Q},\qquad
    \Delta_Q^2=\frac1{R_mN_b}\sum_{r=1}^{R_m} \left\langle\sum_b(M_b^{(r)}-s_is_j)^2\right\rangle_{Q_r},
\label{eq:s_score_self_consistency}
\end{equation}
where the subscript $r$ in $Q_r$ labels the average with fixed measurement outcome $\bm M^{(r)}$.
Because $Q_r$ depends on both $\beta_Q$ and $\Delta_Q$, these equations are coupled and must generally be solved jointly. 
Each defines a constraint in the $(\beta_Q,\Delta_Q)$ plane, whose intersection determines the inferred parameter pair.

At the true parameters, $Q$ becomes the optimal Bayes posterior.
Averaging over records then gives
\begin{equation}
    \E_M[\langle E\rangle_{\rm B}]=\langle E\rangle_\beta,\qquad
    \Delta^2=\frac1{N_b}\E_M\left\langle\sum_b(M_b-s_is_j)^2\right\rangle_{\rm B}.
\label{eq:s_score_true}
\end{equation}
These identities follow directly from the Bayes relation $p_{\beta,\Delta}(\bm M)p_{\beta,\Delta}(\bm s|\bm M)=p_{\beta,\Delta}(\bm s,\bm M)$.

\end{widetext}

\end{document}